\documentclass[aps,prx,twocolumn,groupedaddress,amsmath,amssymb,reprint, superscriptaddress,longbibliography]{revtex4-2}
\usepackage{graphicx}
\usepackage{dcolumn}
\usepackage{bm}
\usepackage{float}
\usepackage[caption=false]{subfig}
\usepackage{hyperref}
\DeclareSymbolFont{matha}{OML}{txmi}{m}{it}
\DeclareMathSymbol{\varv}{\mathord}{matha}{118}
\hypersetup{colorlinks,breaklinks,linkcolor=blue,urlcolor=blue,
anchorcolor=blue,citecolor=blue}
\usepackage{color}
\usepackage{natbib}
\usepackage{multirow}
\usepackage{siunitx}
\usepackage{graphicx}
\usepackage{amsmath}
\usepackage[caption=false]{subfig}
\usepackage{siunitx}
\usepackage{placeins}
\usepackage{color}
\usepackage{standalone}
\usepackage{dcolumn}
\usepackage{bm}
\usepackage{microtype}
\usepackage{etoolbox}
\usepackage{amssymb}
\usepackage{mathrsfs}
\usepackage{accents}
\usepackage[normalem]{ulem}
\usepackage[dvipsnames]{xcolor}
\usepackage[all]{hypcap}
\usepackage{enumerate}
\AtBeginDocument{\usepackage{booktabs}}

\begin{document}

\preprint{APS/123-QED}

\title{Beyond the ordinary acoustoelectric effect: superluminal phenomena in the acoustic realm and phonon-mediated Bloch gain}

 \author{A. Apostolakis} \email[]{Apostolakis@fzu.cz}
 \affiliation{Department of Condensed Matter Theory, Institute of Physics CAS, Na Slovance 1999/2, 182 21 Prague, Czech Republic}
 \affiliation{Department of Physics, Loughborough University, Loughborough LE11 3TU, United Kingdom}
 
 \author{ A. G. Balanov} \email[]{A.Balanov@lboro.ac.uk}  
 \affiliation{Department of Physics, Loughborough University, Loughborough LE11 3TU, United Kingdom}
 \author{F. V.~Kusmartsev}
 \affiliation{Department of Physics, Loughborough University, Loughborough LE11 3TU, United Kingdom}
 \affiliation{College of Art and Science, Khalifa University, PO Box 127788, Abu Dhabi, United Arab Emirates}
 \affiliation{Microsystem and  Terahertz Research Center, Chengdu, 610200, P.R. China}
 \author{K. N.~Alekseev} 
 \affiliation{Department of Physics, Loughborough University, Loughborough LE11 3TU, United Kingdom}
\affiliation{Center for Physical Sciences and Technology, Vilnius LT-10257, Lithuania}

\begin{abstract}
It has been shown that coherent phonons can be used as a potent tool for controlling and enhancing
optoelectronic and  transport properties of nanostructured materials. Recent studies revealed
that interaction of acoustic phonons and fast-moving carriers in semiconductor heterostructures
can be accompanied by electron-phonon instabilities that cause ordinary and induced Cherenkov
effects. However, the development of such instabilities is still poorly understood. Our study shows
that other supersonic phenomena, beyond the Cherenkov instability, are possible for non-equilibrium
charge transport in the miniband semiconductor superlattices (SLs) driven by an acoustic plane wave.
Using semiclassical nonperturbative methods and elements of the bifurcation theory, we find the
conditions for the onset of dynamical instabilities (bifurcations) which are caused by the emission 
of specific SL phonons by supersonic electrons, and their back action on the electrons. Notably, the underlying
radiation mechanism is connected  either to normal or anomalous Doppler effects in full accordance with the Ginzburg-Frank-Tamm theory. The appearance of induced  Doppler effects is also discussed 
in relation to the formation of electron bunches propagating through the spatially periodic structure of the SL.
When the amplitude of the acoustic wave exceeds certain threshold, the dynamical instabilities developed in the system are 
manifested as drift velocity reversals, resonances in sound attenuation and absolute negative mobility. 
We demonstrate that the discovered superluminal Doppler phenomena can be utilized for tunable broadband amplification and generation of GHz-THz electromagnetic waves, which creates a ground for development of novel phononic devices. 

\end{abstract}


\date{\today}

\maketitle

\section{\label{sec:level1}INTRODUCTION}
The mediating role of lattice vibrations (phonons)  in  nanocrystals  has been readily recognized as a fundamental tenet in condensed matter physics. In particular, understanding in depth the transport properties of phonons and their interactions with electrons  is important for enhancing the efficiency of thermoelectric  nanostructures \cite{hu2020machine,maire2017heat}  and  developing acoustic metamaterials  \cite{lu2009phononic,davis2014nanophononic} or novel spectroscopic schemes \cite{mante2015thz,sitters2015acoustic}. In the past few years, there has been an intensive research 
 aimed to study the connection between physics of coherent phonons excitation and nonequilibrium  dynamics in electronic systems.  Prominent examples include   phononic devices based on two-dimensional (2D) materials \cite{greener2018coherent,tamagnone2018ultra,poyser2018high,andersen2019electron} which can exhibit high electron mobility \cite{poyser2018high,andersen2019electron},  high-frequency phonon transducers \cite{huynh2008subterahertz,poyser2015weakly,wilson2018evidence} and quantum structures where coherent acoustic phonons have been generated in the THz-GHz range \cite{greener2018coherent,kent2006acoustic, wang2020ultrafast,  shinokita2016strong}. Recently, there has been increased interest in  amplification of acoustic phonons due to their interaction with fast moving carriers in semiconductor superlattices  (SLs) \cite{shinokita2016strong}. 
These experiments  revealed the importance of electron-phonon instabilities where ordinary and induced Cherenkov effects come into play. 
In general, the Cherenkov effect is the well-know phenomenon encountered in the electrodynamics when a charged particle passes through a dielectric medium at speed greater than the phase velocity of light in the medium.  We likewise note new research efforts devoted 
 to the radiation dynamics of the superluminal particle \cite{kaminer2017spectrally,yang2018maximal,roques2018nonperturbative,shi2018superlight}. Interesting examples include  further developments  in Ginzburg and Frank theory \cite{ginzburg1947doppler,Ginzburg60-UFN} describing the Doppler effects \cite{shi2018superlight,lin2019normal},  Smith-Purcell radiation in plasmonic crystals \cite{kaminer2017spectrally} and nonperturbative generalization
of Cherenkov radiation \cite{roques2018nonperturbative}. In the acoustic realm though, Cherenkov emission can be induced when the average electron velocity ($v_d$) in the presence of a static field  moves faster than the speed of sound $(v_s)$. This supersonic condition ($v_d>v_s$) is well satisfied in an electrically biased SL \cite{shinokita2016strong} where the propagating acoustic wave interacts with the electrically driven electron current.
\begin{figure*}[ht]
   \includegraphics[width=0.8\linewidth]{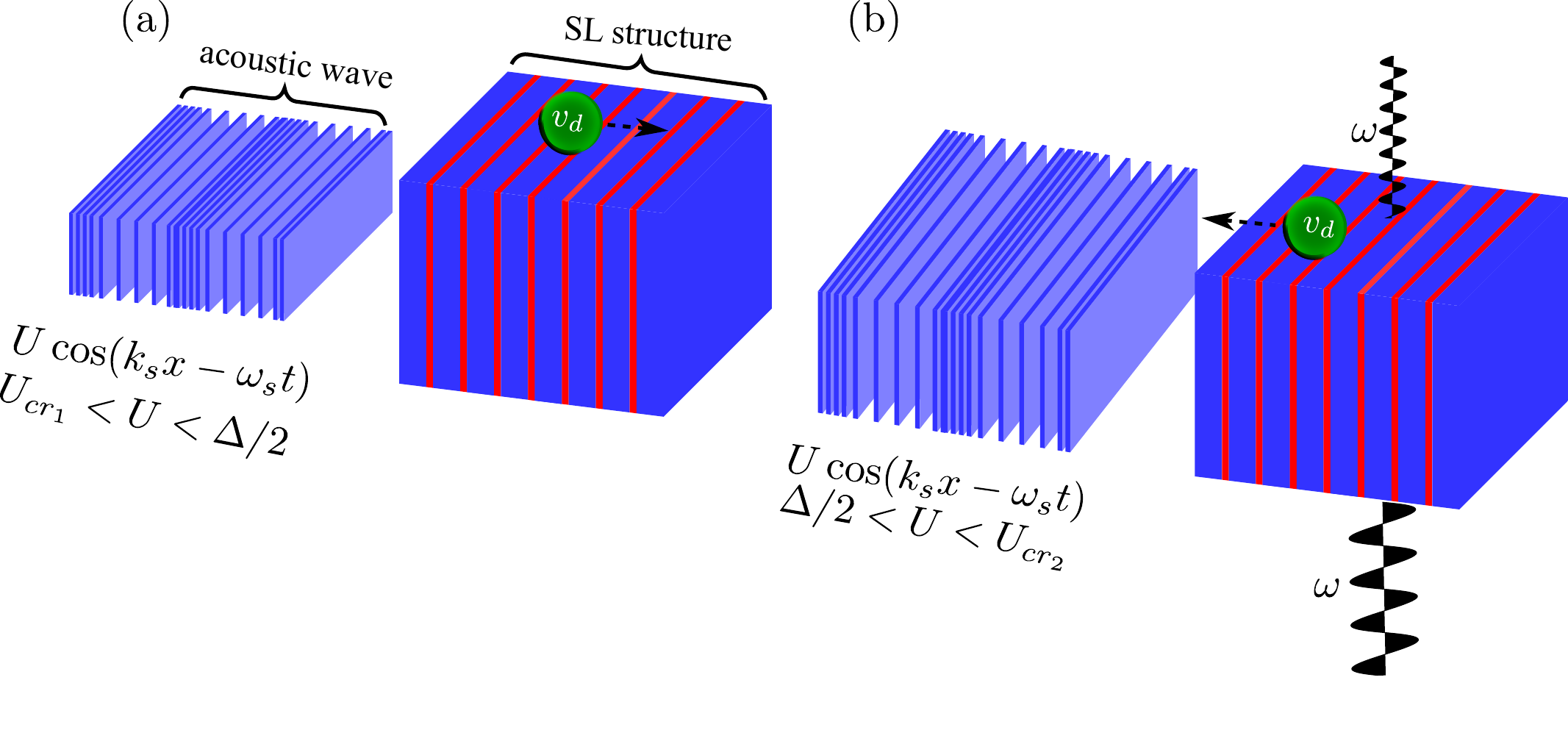}
   \caption{Schematic diagram of novel electroacoustical effects beyond the ordinary charge drag.
A strong  acoustic plane wave of the energy amplitude $U$ propagates at $v_s=\omega_s/k_s$  along the growth direction ($x$--axis) of a superlattice (SL) with a  miniband width of $\Delta$ and  lattice period $d$. (a) For 
$U<\Delta/2$ charges are always transported along the direction of the wave propagation. However, bunches of electrons begin to oscillate strongly at $U=U_{cr_1}$, which is slightly below $\Delta/2$ 
 [see Eq. (\ref{eq:phys-bifur})], and then drift of the bunches quickly slows down with an increase of $U$.  
(b) For $U>\Delta/2$  
the electronic bunches move opposite to the propagating wave, and their backward drift becomes
fastest at $U_{cr_2}$, which value is slightly above $\Delta/2$  
 [Eq. (\ref{eq:phys-bifur})]. 
Our analytical selection rules [Eq. (\ref{eq:phys-bifur})] indicate that the critical amplitudes $U_{cr_1}$ and $U_{cr_2}$ 
mark respectively superluminal anomalous and superluminal backward Doppler regimes [Eq. (\ref{eq:doe})] of emission  and absorption of specific phonons [Eq. (\ref{eq:phys-bifur})] by the electrons. Additionally, at the wave amplitude, $U_{cr_2}$,
an interaction of the electrons with a high-frequency ($\omega$) electromagnetic wave can result in its amplification. } 
\label{fig1}
\end{figure*} 
The aforementioned study  further confirmed that Cherenkov effects and possibly other more exotic electromagnetic phenomena might have acoustic counterparts. It was only recently, when semiclassical studies and  quantum-mechanical simulations \cite{greenaway2010using,apostolakis2017nonlinear,wang2020ultrafast} predicted that under the action of a strong acoustic wave, the propagating deformation potential can induce quasiperiodic Bloch oscillations of miniband electrons. The rise of these complex Bloch oscillations was  linked to global instabilities triggered with an increase of the wave amplitude, which serves as control parameter \cite{apostolakis2017nonlinear}.  
 Remarkably, the development of these instabilities clearly distinguishes the case of SL from the ordinary electroacoustic effect in bulk semiconductors with a quadratic band \cite{Parmenter53,gulyaev2005acoustoelectronics}. 
\\In this work, we study supersonic phenomena that should  appear when a coherent acoustic stimulus induces high-frequency electron dynamics in semiconductor superlattice. In particular, we theoretically  consider acoustically driven miniband electrons which tunnel through the SL periodic potential  under the action of a longitudinal strain wave with energy amplitude $U$, as shown in Fig. \ref{fig1}(a). 
 In this case, the acoustic 
  stimulus interacts with the electrons by means of a deformation  potential.
 We demonstrate that this interaction and 
 the  induced miniband transport are controlled by the laws  of physics 
  typical for superluminal particles with internal degrees of freedom \cite{Nezlin76}.
{ In our case, the internal degrees of freedom are associated with the propagating strain-induced deformation potential. Analysis of the semiclassical phase space dynamics in the reference frame moving at $v_s$ allowed to identify specific dynamical instabilities (bifurcations) which are developed for certain critical values of the strain-induced  potential  $U=U_{cr_n}$ ($n=1,2...$, $U_{cr_1} < U_{cr_2} < ...$). Tuning $U$ to these values induces the formation of new bound states between the electrons and the phonons.
We found out that the phonon-assisted 
 transport involving  those bound states are, in essence, a direct generalization of the normal and anomalous Doppler  effects discussed in the radiation theories introduced by Ginzburg and Frank for supreluminal photons \cite{Ginzburg60-UFN}.  Our model parameters 
correspond to the regime 
related to the recently discovered superlight inverse Doppler effect \cite{shi2018superlight}. Moreover, our results provide compelling arguments for the long-standing idea introduced by  Tamm \cite{Tamm60} that the characteristics of the Ginzburg-Frank radiation theory can be implemented to supersonic acoustical systems. The emitted phonons which are involved in the radiation processes are found to behave in a manner similar to a Smith-Purcell (SP)  photons \cite{SmithPurcell53,Bolotovskii68}. 
\\ Our study showed
that stimulated emission creates 
electron bunches 
which move along the spatially periodic structure of the SL. For both normal and anomalous Doppler behavior, the electron bunching 
effects on directed charge transport appear to be robust even when we take into account scattering processes.  
Remarkably,  the involvement of  the anomalous Doppler effects for  $U$ $<\Delta/2$ 
 decelerates the single electron with the overall drift velocity remaining positive. Here  $\Delta$ is the first minband width, within which a charge tunneling transport is assumed.
 For $U>\Delta/2$, the reversal of the drift results from the emergence of both anomalous and  conventional normal Doppler shifts. 
\\ Signatures of the Doppler effects are also found in attenuation of of the acoustic wave, which is characterized by the calculated absorption coefficient. 
We  also demonstrate  a possibility of absolute negative mobility (ANM) for a SL structure in combined  dc bias and acoustic wave drive with sufficient large amplitude to induce superluminal Doppler effects. Notably, ANM has already attracted great research interest in connection to spontaneous generation of large static fields \cite{alekseev1998spontaneous}  and high-frequency stimulated emission of photons \cite{keay1995dynamic} when the superlattice structure is driven by intense terahertz electric fields.\\ Our analysis shows that the discovered phenomena can be exploited practically in 
 schemes for tunable broadband amplification and generation of GHz-THz electromagnetic waves. Note that  SLs under moderate electric fields  have been already 
shown as a system to provide optical gain due to the Bloch oscillations in the presence of weak dissipation \cite{Ktitorov72,ignatov1976nonlinear}. This prediction was based on semiclassical arguments  and coined as Bloch gain, raising the possibility of inversionless lasing  in  a dc biased SL.  The use of phononic
waves though, opens new opportunities to enhance the performance of superlattice
oscillators. Here  we put forward a scheme for the broadband  amplification of THz radiation in acoustically driven superlattice similar to the Bloch gain in a electrically biased SL.\\
This paper is organized as follows. In Sec. \ref{sec:level2}, we employ  elements of the bifurcation theory and  we consider the conditions for which the superluminal nature of electron kinematics arises. Therefore, we identify a class of Cherenkov and Doppler resonances which are directly connected to global bifurcations developing with an increase of the wave wave amplitude. In Sec. \ref{sec:level2a}, we focus on the stimulated emission of the SL-phonons for choices of the wave amplitude between the analytical bifurcations points which is confirmed by numerical calculations determining the kinetic behavior of a classical ensemble of particles.   In Sec. \ref{sec:level3}  a non-perturbative solution of  the Boltzmann transport equation is followed to examine the main transport characteristics in SL at the presence of scattering.   We discuss the feasibility of ANM, for an electrically biased SL in Sec.  \ref{sec:level4} and the broadband amplification of an electromagnetic (EM)  wave for an acoustically pumped SL in Sec. \ref{sec:level5}. We conclude
in Sec. \ref{sec:level6} with a few remarks.

\section{\label{sec:level2} Superluminal phenomena in the realm of acoustoelectric interactions}

This section discusses  the  nature of the acoustoelectric effects in SLs.  The nonlinear analysis of the acoustoelectric instabilities provides a fruitful insight for the absorption of the acoustic wave, its scattering by miniband electrons and the related quantum processes. This would allow us to show that a wide range of superluminal phenomena have distinct acoustic counterparts.\\ We consider a longitudinal acoustic wave that propagates in the direction of the superlattice axis ($x$) [Fig.~\ref{fig1}(a)] generating a position and time-dependent potential energy  given by  \cite{Kazarinov63,greenaway2010using}
\begin{equation}
V(x,t)=-U\sin[(k_s(x+x_i) -\omega_s t)],
\label{eq:penergy}
\end{equation}
where $U=D \epsilon$ is defined by  the deformation potential constant $D$  and the strain magnitude  $\epsilon$.
The   displacement $x_i$ defines the initial phase of the driving wave, $\omega_s$ is the wave frequency, $v_s$ is the speed of sound in the materials of SL,  $k_s=\omega_s/v_s$ is the wave number, and $t$ is time. In the simplest tight-binding scheme, it sufficient to describe the dispersion of the first SL miniband as 
\cite{esaki1970superlattice}
\begin{equation}
\mathcal{E}(p_x)=\dfrac{\Delta}{2}\left[1-\cos\left(\dfrac{p_x d}{\hbar}\right)\right],
\label{eq:disenergy}
\end{equation}
where $\mathcal{E}$ is the energy of an electron with quasi-momentum $p_x$, $\Delta$ is the miniband width, and $d$ is the SL  period.
In general, the dispersion relation should also include a contribution of the lateral (along the superlattice layers) motion in the form $\mathcal{E}_\parallel(p_y, p_z)=p^2_y/2m^*+p^2_z/2m^*$, where $m^*$ is the effective electron mass. However, without account of anisotropic scattering the electron motion in lateral directions is independent from electron tunnelling along the superlattice axis, and therefore the most interesting physical effects can be described by the semiclassical  Hamiltonian
\begin{equation}
H(x, p_x)=\mathcal{E} (p_x)+V(x,t).
\label{eq:hamiltonian1}
\end{equation}
\begin{figure}[t]
   \includegraphics[scale=1]{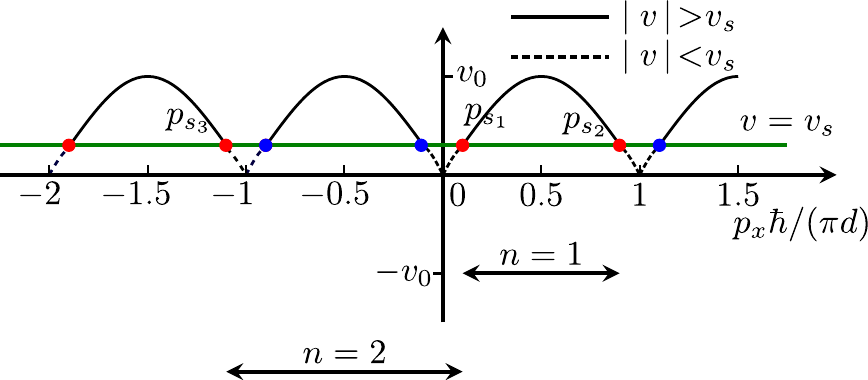}
   \caption{ The modulus of electron velocity $|v(p_x)|$ [Eq. (\ref{xdot})] represented by dashed lines corresponding to subsonic charge carriers and  solid curves  to supersonic electrons that belong to the active zone  and therefore they can emit or absorb phonons while making transitions within the miniband.   The green horizontal line connects  the hyperbolic points $p_{s_i}$ [Eq. (\ref{eq:stp})], with $i=1, 2, 3$, denoting the condition for absorption of the acoustic wave for small $U$ [see Eq. (\ref{eq:all-usfp_tilde})]. The  arrows ($n=1$) and ($n=2$) indicate which hyperbolic points are connected in order to satisfy the selection rules described by Eq. (\ref{eq:phys-bifur}).}
\label{fig2} 
   \end{figure}
It is easy to see from this Hamiltonian, the electron velocity, $\partial \mathcal{E}/\partial p_x $ and  the force acting on the electrons, $-\partial H/\partial x $, due to the propagating deformation potential  (\ref{eq:penergy}), which read respectively
\begin{subequations}
\label{eq:all-dot}
\begin{eqnarray}
v&=&\frac{\partial \mathcal{E}}{\partial p_x }=v_0\sin\left(\frac{p_xd}{\hbar}\right),  \label{xdot}\\
\frac{dp_x}{dt}&=&-\frac{\partial H}{\partial x } =k_sU\cos[(k_s(x+x_i) -\omega_s t)], \label{pdot}
\end{eqnarray}
\end{subequations}
with $v_0=\Delta d/(2\hbar)$ being the maximal miniband velocity.  It  has been demonstrated that this simple model well describes the experiments on charge transport in SL driven by a train of picosecond strain pulses \cite{poyser2015weakly,wang2020ultrafast}.  We  consider mainly two different sets of parameters which are summarized in  Table \ref{tab:firstable} corresponding to  $\textrm{GaAs}/\textrm{Al}_x\textrm{Ga}_{1-x}\textrm{As}$ SLs used in recent experiments  \cite{fowler2008semiconductor,shinokita2016strong}. Our model (\ref{eq:all-dot}) though  has comfortably allowed us to calculate similar results for a wide range of SL parameters.
\begin{table}[H]
\centering
  \caption{Basic parameters of superlattices studied.  The frequency of sound  is assumed $\omega_s=4\times 10^{11}$ rad/s with $\omega_s \sim 0.1 /\tau$ and $k_s\sim 1/d$, unless stated otherwise. Further details are given in appendix \ref{App5}.   }
    \begin{ruledtabular}
   \begin{tabular}{c@{\quad} c@{\quad} c@{\quad} c@{\quad} c@{\quad} c@{\quad} c@{\quad} c@{\quad}}
    $\Delta$ (meV) & $d$ (nm) & $\tau$ (fs) & $v_s $ (m/s)& $D$ (eV) &  \\
    {\tiny miniband width} &{\tiny lattice period} & {\tiny scattering time} & {\tiny velocity  } & {\tiny Deformation }  \\ {} &{} & {} & {\tiny of sound} & {\tiny potential} \\
    \midrule
    7   & 12.5  &  \multirow{2}{*}{250} & \multirow{2}{*}{5000} & \multirow{2}{*}{10}\\
    20  & 11.4 &   &  &  &  \\

  \end{tabular}
  \label{tab:firstable}
    \end{ruledtabular}
\end{table}
\textit{Efficient interaction of electrons with phonons.}--- We turn now to the condition for efficient interaction of the band electrons with acoustic phonons  that was initially introduced in the theory of ultrasound absorption in metals: 
In the case the mean free path of electrons is large, the absorption of the sound wave can be considered as scattering of an aggregate of phonons  by those electrons whose velocity in the direction of the acoustical wave vector is equal to the phase velocity of sound \cite{galperin1969giant,abrikosov2017fundamentals}. Indeed, for $v>v_s$ the conservation laws should dictate that for the absorption of a phonon
\begin{subequations}
\label{eq:absorption}
\begin{eqnarray}
p_f&=&p_i+\hbar q ,\label{eq:mom_con} \\
\mathcal{E}(p_f)&=&\mathcal{E}(p_i) +\hbar \omega_q, \label{eq:en_con}
\end{eqnarray}
\end{subequations}
where $p_f$ ($p_i$) stands for the final (initial) momentum of the electron, and $\hbar q>0$  is the quasimomentum of the absorbed phonon.  For generality we assume here that the phonon frequency $\omega_q$ can be different from the frequency of the propagating deformation potential (\ref{eq:penergy}). Combining Eqs. (\ref{eq:mom_con}), (\ref{eq:en_con}) we obtain 
$\mathcal{E}(p_i+\hbar q)=\mathcal{E}(p_i) +\hbar \omega_q $. Since the momentum exchanged between the low-frequency phonons  and band electrons is very small, $\hbar q\ll p_i$, the energy then $\mathcal{E}(p_i+\hbar q)$ can be expanded into Taylor series as
\begin{equation}
\mathcal{E}(p_f)
\approx
\mathcal{E}(p_i)+\hbar v(p_f)q.
\label{eq:energyexp}
\end{equation}
By comparing Eq. (\ref{eq:energyexp}) and Eq. (\ref{eq:en_con}) we have
\begin{equation}
   v\approx v_s=\omega_q/q,  
\label{eq:stp}
\end{equation}
which corresponds to Cherenkov effect \cite{galperin1969giant,ashley1965phonon,rivera2020light}. It can be shown that the same condition (\ref{eq:stp}) needs to be satisfied in order an electron emits a phonon with a small quasi-momentum $\hbar q$. 
In this derivation it was implicitly assumed that influence of the  electron potential energy in scattering events is negligible, and also that the electrons oscillate far from edges of the first Brillouin zone. Now we are going to show that the resonant condition (\ref{eq:stp}) naturally arises in the analysis of the fixed points of the dynamical system (\ref{eq:hamiltonian1}). 
\\ \textit{Fixed points and their physical meaning.}--- We first make a canonic transformation to the moving with the velocity $v_s$ reference frame, for which new Hamiltonian, electron kinetic energy and coordinate take the forms
$$
H'=\mathcal{E}'(p_x)+V(x'),  \quad
\mathcal{E}'(p_x)=\mathcal{E}(p_x)-v_s p_x, 
$$
$$
x'(t)=x(t)+x_0-v_s t,
$$
where the expressions for  $V(x)$ and $\mathcal{E}(p_x)$  are still given by Eqs. (\ref{eq:penergy}) and (\ref{eq:disenergy}). While the Hamiltonian $H'$ becomes time-independent, the electron momentum $p_x$ is unchanged under the canonic transformation.
}
 In Appendix \ref{App0}, we  showed that the fixed points of this autonomous dynamic system are 
\begin{subequations}
\label{eq:all-usfp_tilde}
\begin{eqnarray}
x'&=&\frac{\pi}{2 k_s} +\frac{m\pi}{k_s},  \label{eq:unsfp_xtilde}\\
p_x&=&(-1)^{l}\frac{\hbar}{d} \sin^{-1}\left(\frac{v_s}{v_0}\right) +l\frac{\hbar\pi}{d}, \label{eq:unsfp_ptilde}
\end{eqnarray}
\end{subequations}
where $m$ and $l$ are arbitrary integer numbers.\\  To understand physical meaning of the fixed points we first consider the case of
small $U$ when $p_x(t)$ oscillations are well confined within the first Brillouin zone $|p_x d/\hbar|< \pi$, evoking thereby only the conventional acoustoelectric response. Importantly, Eq.~(\ref{eq:unsfp_ptilde}) for the $p-$components originates from the condition $v(p_x)=v_s$ [cf. Eq.~(\ref{eq:fpx})], 
which is identical to Eq.~(\ref{eq:stp}).
Therefore,  Cherenkov absorption or emission of phonons can arise in the proximity of the stationary points. Figure \ref{fig2} demonstrates schematically the modulus of electron velocity in the momentum subspace  and the positions therein of the hyperbolic  points (red circles):  $p_{s1}=(\hbar/d)\sin^{-1}(v_s/v_0)$, $p_{s2}=\pi d/\hbar-(\hbar/d)\sin^{-1}(v_s/v_0)$  and $p_{s3}=-\pi \hbar/d-(\hbar/d)\sin^{-1}(v_s/v_0)$  which are connected by a  horizontal line  corresponding to the condition [see Eq. (\ref{eq:stp})] for the direct absorption of phonons by electrons. More detailed consideration shows that both forward and backward Cherenkov \cite{chen2011flipping,rivera2020light} effects can exist for supersonic miniband electrons. For extended discussion see Appendix \ref{App0}.\\ 
The back action of such Cherenkov phonons on the miniband electrons is definitely small. 
In what follows, we will discuss an opposite situation of significant changes in the transport of the miniband electrons caused by their interactions with such phonons that are able to carry large momenta and therefore able to exert energy transitions both in the miniband (\ref{eq:disenergy}) and in the potential wells formed my the acoustic wave (\ref{eq:penergy}). Of course, this will require relatively large amplitudes of the wave $U\gtrsim \Delta/2$, when effects of the electron potential energy cannot be ignored any more.\\
 \textit{The emergence of  Doppler effects.}---
Next, we will reveal  superluminal mechanisms 
  and the related instabilities (bifurcations) that are developed for larger  $U$.
 A series of global bifurcations with the increase of $U$ have been analytically found previously  for the model described by Eq. \ref{eq:all-dot} \cite{apostolakis2017nonlinear}.
 It has been shown that they are associated with the reconnection of the hyperbolic points  [see Eq. \ref{eq:all-usfp_tilde}]  by separatrices in the phase space of the dynamical system (\ref{eq:all-prime}) at  the following critical values of $U$:
\begin{subequations}
\label{eq:u_cr}
\begin{eqnarray}
U_{cr_n}&=&\frac{\Delta}{2}\sqrt{1-\left(\frac{v_s}{v_0}\right)^2}+
\frac{\hbar v_s q_n}{2} \quad (n\geq 1), \label{eq:u_cr-itself} \\
q_n&=&\frac{2}{d}\sin^{-1}\left(\frac{v_s}{v_0}\right)+\left(2n-3\right)\frac{\pi}{d}. \label{eq:u_cr-q}
\end{eqnarray}
\end{subequations}
where $n$ is an integer designating the order of the bifurcation. 
 From physical point of view, Eqs. (\ref{eq:u_cr}) imply multi-phonon resonances in a transition scattering of the sound wave by electrons \cite{Ginzburg_bristol} placed in the SL periodic potential. 
 Specifically, the second term in Eq. (\ref{eq:u_cr-itself}) defines an acoustic phonon quantum $\hbar\omega(q_n)=\hbar v_s q_n$ with the effective wave number  $q_n$ (\ref{eq:u_cr-q}), while the first term in Eq. (\ref{eq:u_cr-itself}) is proportional to the width of the active zone in energy band, which will be discussed later.  The emission and absorption of phonons with wavenumber $q_n$ [Eq. (\ref{eq:u_cr-q})], however,  at  critical (bifurcation) points result in both normal  ( $n=1$) and Umklapp ($n\geq 2$) scattering transitions of electrons in momentum space. As can be seen in Fig.  \ref{fig2},  the two hyperbolic points ($p_{s_1}$, $p_{s_2}$) which are involved in the first bifurcation (Fig. \ref{fig13}) with critical value
 \begin{equation}
U_{cr_1}=\frac{\frac{\Delta}{2}[\cos (p_{s_1}d/\hbar)-\cos (p_{s_2} d/\hbar)]+v_s p_{s_1}-v_s p_{s_2}}{\sin(k_s x_{s_2})-\sin(k_s x_{s_1})}, \label{eq:u_cr-qsp1}
\end{equation}
 lie within the first 
 BZ resulting in an electron normal scattering transition in $p_x$--space.   In contrast, an electron Umklapp scattering stems from the second bifurcation with critical value 
  \begin{equation}
U_{cr_2}=\frac{\frac{\Delta}{2}[\cos (p_{s_2}d/\hbar)-\cos (p_{s_3} d/\hbar)]+v_s p_{s_2}-v_s p_{s_3}}{\sin(k_s x_{s_3})-\sin(k_s x_{s_2})}, \label{eq:u_cr-qsp2}
\end{equation}
 in which  the separatrix is reconnected (see Fig. \ref{fig14}) through ($p_{s_1}$, $p_{s_3}$) with $p_{s_3}$ being located outside the first  BZ.
 Note that Eqs. (\ref{eq:u_cr}) for $n$=1, 2 can be  derived from  Eq. (\ref{eq:u_cr-qsp1}) and (\ref{eq:u_cr-qsp2}), respectively, by just rearranging  their terms. 
\\ \textit{The active zone region.}---The stationary points of hyperbolic type  ($x_s$, $p_s$) define  intervals of supersonic electron motion in the momentum subspace  $p_x$. Within these intervals the momentary electron velocity $v(p_x)$ can exceed the sound velocity $v_s$. In detail,  the pairs of resonant points ($p_{s1}, p_{s_2}$) and ($p_{s1}, p_{s_3}$)  determine intervals within the first Brillouin zone, $0<p_x<\pi \hbar/d$ and  $-\pi \hbar/d<p_x<0$ respectively, in which  the supersonic motion (with the velocity in Fig. \ref{fig2} represented by black solid lines) takes place.  It follows that  $v(p_x)>v_s$ holds for an energy interval that is centered at the middle of the energy band 
$\mathcal{E}=\Delta/2$,  and has the width
\begin{equation}
\delta \mathcal{E}_{v>v_s}=\Delta \left[ 1-\left(\frac{v_s}{v_0}\right)^2\right]^{1/2}. \label{eq:deltaE-gen}
\end{equation}
Supersonic electrons belonging to the ``active zone'' defined by  (\ref{eq:deltaE-gen}) can emit phonons  while following the phase trajectories defined by  Eq. (\ref{eq:pp}).
When the maximal miniband velocity $v_0$ is approaching the speed of sound $v_s$,
the active zone becomes narrow $\delta \mathcal{E}_{v>v_s}\rightarrow 0$. \\
 \textit{Hypersonic limit.}---We now consider  the implications of the opposite limit $v_0/v_s\gg 1$, which  is well satisfied for typical semiconductor SLs [see Table \ref{tab:secondtable}] and where the superluminal physics become more transparent. In this case, the active zone practically coincides with the entire miniband, $\delta \mathcal{E}_{v>v_s}\rightarrow\Delta$ and therefore   practically eliminating the subsonic regions (black dashed lines) between blue and red circles in  Fig. \ref{fig2}; see discussion in appendix \ref{App5}. Physically, it guarantees that the miniband electron speed $v=v_0|\sin (p_x d/\hbar)|$ exceeds the speed of sound $v_s$ almost for any momentum $p_x$. 
Additionally,  the hyperbolic points (\ref{eq:unsfp_ptilde}), involved in the derivation of Eq. (\ref{eq:u_cr}), are located 
at $p_s\approx0$ and very close to the boundaries of the first and successive BZs.
 In what follows, we will mainly focus on the first two major bifurcations, for which 
\begin{equation}
U_{cr_{1,2}}=\frac{\Delta}{2} \mp \frac{\hbar\omega_q}{2}, \hspace{0.1cm} \omega_q=\pi v_s/d,
\label{eq:phys-bifur}   
\end{equation}
 since their effects play a fundamental role, as we will see, in the electron kinetics. 
It is easy to see that the relative contribution of the terms $\Delta/2$ and $\hbar\omega_q/2$ to the value $U_{cr}$ in equation~(\ref{eq:phys-bifur}) is defined  by the ratio $v_0/v_s$.
 Therefore, in the physically interesting limit $v_0/v_s \gg 1$ the critical wave amplitude $U_{cr}$ asymptotically approaches  the value 
 $U_0=\Delta/2$, i.e.  the half of the miniband width. The limit $v_0/v_s\rightarrow\infty$ itself can be reached either by increasing $\Delta$, or by slowing down the speed of sound  $v_s\rightarrow 0$. The later demonstrates that the appearance of the quantum in  criterion (\ref{eq:phys-bifur})  is directly related to the propagation effects.
\begin{figure}[t]
\includegraphics[scale=0.7]{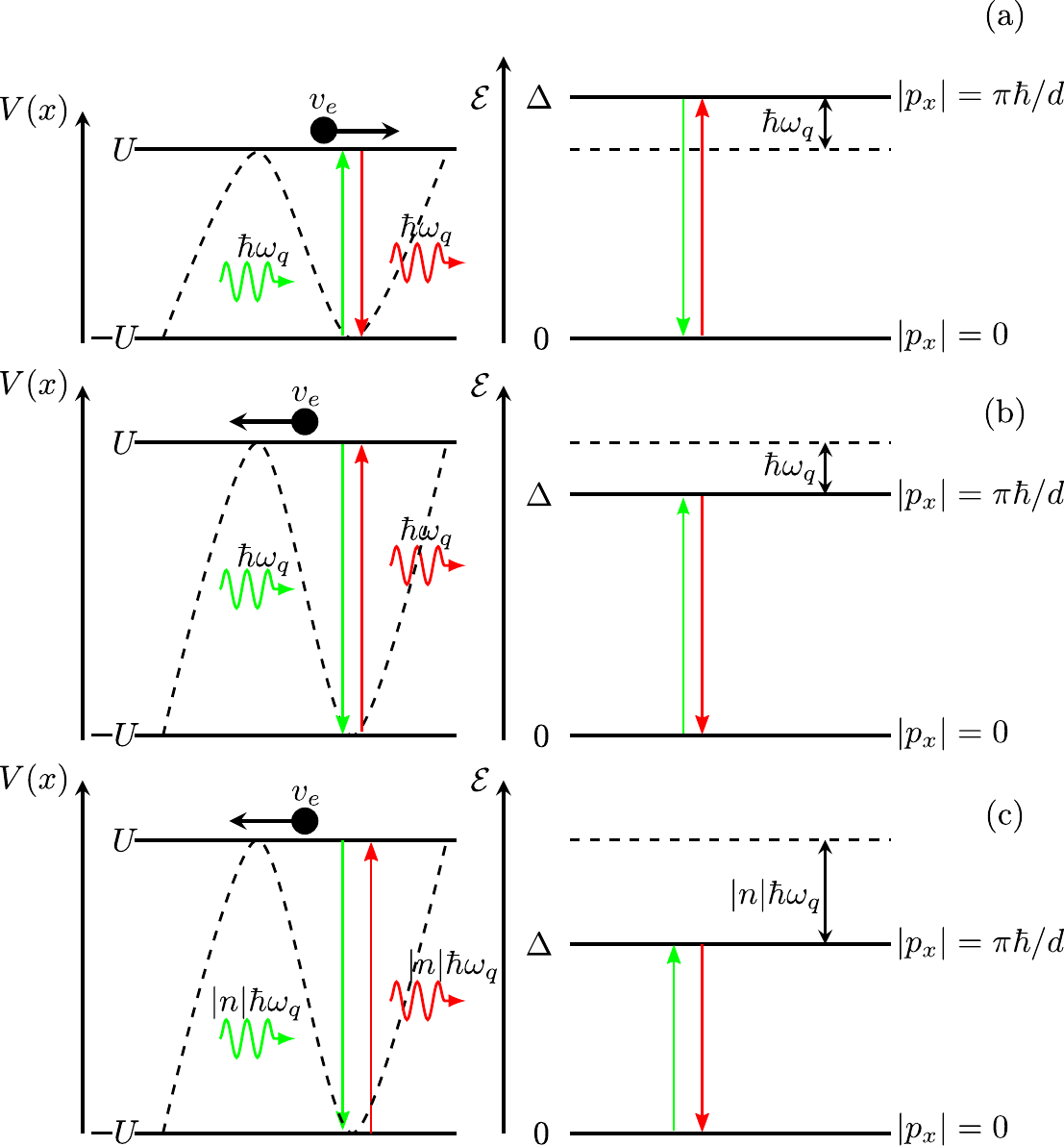}
   \caption{Diagrams of electron transitions within the potential of the acoustic wave (left panel) and within the energy miniband (right panel) manifested by anomalous and normal Doppler effects corresponding to the bifurcations in increasing order. (a) The electron trapped by the wave $V(x)$ can either absorb the phonon $\hbar\omega_q$ and make a transition from the top of the potential $V_{min}=U$ to its bottom $V_{max}=-U$ or follow the reverse process by emitting a quantum.  (b) The electron recoils, while emitting a phonon, leading to a normal Doppler shift of the transition (green arrow). Alternatively, the electron experiences an excitation (red arrow) assisted by absorption of $\hbar\omega_q$. (c) One-step multiphonon transitions involving odd number of normal Doppler shifts.The green lines describe emission processes  where the red lines absorption ones. }
	


\label{fig3}
\end{figure}
\\To get a deeper insight into the physical meaning of Eq. (\ref{eq:phys-bifur}),
consider an electron that absorbs or emits a phonon with the quasi-momentum $\hbar q=\hbar 2\pi/\lambda_q=h/2d$. 
As a result of the radiation/absorption act, the electron momentum becomes $p_f=p_i +\hbar q$, where $p_f$ ($p_i$) stands for the final (initial) momentum and $q$ is positive or negative depended on whether we have absorption or emission processes. Next, a variation of the electron kinetic energy in the  moving reference frame of the acoustic wave  is 
\begin{equation}
\delta \mathcal{E}'=\delta \mathcal{E}-v_s\delta p,
\label{eq:emf}   
\end{equation}
 where $\delta \mathcal{E}=\mathcal{E}(p_f)-\mathcal{E}(p_i)$ and $\delta p=p_f-p_i$. Assuming the electron is initially at the center of Brillouin zone $p_i=0$, it is  easy to find 
both the variation of the electron kinetic energy  $\delta \mathcal{E}=\Delta$ and the variation of its momentum $\delta p=\hbar\omega_q/v_s$. Finally, equating $\delta \mathcal{E}'$ to the corresponding variation of the electron potential energy $\delta V=V_f-V_i$, we get
the condition of energy conservation in the moving frame as  
\begin{equation}
\delta V=-\Delta\pm\hbar\omega_q.
\label{eq:enegy-conserv}   
\end{equation}
Now consider an electron transition from the top 
of the potential $V(x')$  at $x'_s=-\pi/(2 k_s)$ to its bottom  at $x'_s=\pi/(2 k_s)$, for which $\delta V=-U-(+U)=-2U$. Substituting this $\delta V$ in (\ref{eq:enegy-conserv}) we obtain equation~(\ref{eq:phys-bifur}).
Therefore, the values of the critical amplitudes $U_{cr_{1,2}}$ follow from a delicate energy balance in distribution of the acoustic wave energy ($2U$) between an excitation of the electron within the miniband ($\mathcal{E}_0=0\rightarrow \mathcal{E}_f=\Delta$) and absorption   or emission  of the quantum $\hbar\omega_q$. Since we have assumed that $p_i=0$ (lattice temperature close to zero), the first bifurcation  is related to an absorption process in order to satisfy simultaneously Eqs. (\ref{eq:phys-bifur})--(\ref{eq:enegy-conserv}). On the other hand, the same initial condition results in emission of a phonon after the second bifurcation.
While the energy of the quantum is relatively small $\hbar\omega_q/\Delta\ll 1$, it brings a large momentum  inversely proportional to the lattice period $d$
 [see Eq. (\ref{eq:u_cr-q}) for $n=1$]. Thereby this inelastic scattering event is able to kick  the electron from the bottom of the miniband directly to its upper edge, giving rise to the the electron Bragg reflections. 
The condition of energy conservation (\ref{eq:enegy-conserv}) in the moving frame can be rewritten in the alternative form
\begin{equation}
 -\hbar\omega_0=\hbar(-q|v_e|\pm \omega_q),
 \label{eq:doe}
 \end{equation}
where $|v_e|=2 v_0/\pi$ is the effective electron speed and $\omega_0=|\delta V|/\hbar$ is the resonant frequency  which corresponds to the source excitation energy $\delta V$. 
\\ 
Remarkably,  equation (\ref{eq:doe}) 
describes both  the normal and anomalous Doppler effects, depending on the sign in from of $\omega_q$. Namely, entrance of $+\omega_q$ corresponds to anomalous Doppler effect, while $-\omega_q$ implies a normal Doppler effect.
\\ In our system the first resonance  (at $U=$ $U_{cr_1}$) is associated with the anomalous Doppler effect. The anomalous character of the process is  reflected in the fact that despite an electron absorbs the phonon $\hbar\omega_q$ (red wavy arrow), it
makes transition from the top  of the potential to its bottom (red vertical arrow) as illustrated in the left panel of  Fig. \ref{fig3}(a).  Thus, the absorption   of a quantum happens at the expense of potential energy, whereas the electron is excited at the upper edge of the miniband as indicated by the red arrow in the right panel of Fig. \ref{fig3}(a).  Note that the SL parameters considered in this work can also operate in the regime where new phenomena such as the superlight inverse Doppler effect \cite{shi2018superlight}  are possible and thus $|\delta V|>\hbar \omega_q$ whereas the corresponding Doppler frequency shift has $\Delta \omega=\omega_q-\omega_0<$0. Regardless of the anomalous radiation act,  just after the first bifurcation the miniband electron (black circle with the right-handed arrow) continues to move in the positive direction with effective particle velocity, $v_e$. For detailed discussion of electron's kinetic behavior see the following subsection and Sec. \ref{sec:level3}.  In similar way one could also analyze the radiating processes far from the thermal  equilibrium (at high temperature limit)  by considering that the electron could initially sit at the edge of Brillouin zone ($p_i$=$\pi \hbar/d$)  and subsequently inelastically scattered to center of the first Brillouin zone at  $p_i$=0.  This process is depicted by the green vertical arrows of Fig. \ref{fig3}(a). On these terms, the anomalous character of the process would lie in electron emitting the phonon $\hbar \omega_q$ (green wavy arrow) and making a transition  from the bottom of the potential to its top.  Here, the emission of a quantum would happen instead at the expense of the  kinetic energy. In this sense our model can  accommodate a further generalization of the Ginsburg-Frank-Tamm theory \cite{Tamm60}.\\ After the second bifurcation (at $U=$ $U_{cr_2}$), which corresponds to a normal Doppler effect, the realization of electronic transition from the top to the bottom of the potential [green arrow, left panel Fig. \ref{fig3}(b)] is conventionally accompanied by an emission of a phonon. That, however, makes the electron to be excited [green arrow, right panel Fig. \ref{fig3}(b)] within the miniband.  Notably, such a transition  can be associated with the backward Cherenkov radiation \cite{lin2019normal,chen2011flipping}, where the electron (black circle with left-handed arrow) moves in the opposite direction with respect to  the emitted phonon and the associated Doppler-type frequency shift has $\Delta\omega<0$. On the contrary,  for an arbitrary system such as Larmor oscillator \cite{gintsburg1962anomalous}, the  conventional normal Doppler effect holds $\omega_0=\omega_q-q|v_e|$  with $\Delta\omega>0$. The appearance of the Cherenkov effect in periodic photonic structures has demonstrated inherit connections between electron-photon interactions manifested by the Smith-Purcell effect and the Cherenkov effect \cite{luo2003cerenkov}. Similarly, here is the SL  which can a play a role of an effective diffraction grating for Smith-Purcell-type phonons.  We can identify  that resonances (\ref{eq:phys-bifur}) describe the SP mechanism where the emission spectrum  is determined only by the period of the SL sample ($\omega_q\propto 1/d$)  which constitutes an effective diffraction grating. The emitted radiation at wave number $q$ (cf. Eq.~(2) in \cite{Bolotovskii68}) taken in the limits $\beta=v/v_s\gg 1$  is reduced to a form containing the characteristic Doppler factor so that the angle of emission is either $\theta=0$  or $\theta=\pi$. Finally, it is possible to examine the phonon absorption (emission) processes for higher order bifurcations  at $U_{cr_n}$, $n\geq3$, see  Eq. (\ref{eq:u_cr-itself}).
For example, a three-phonon process can occur if the electron is inelastically scattered by a strong  sound wave (at $U=U_{cr_3}$) which allows to perform a radiating transition from the bottom to the top of the miniband [see the right panel of Fig. \ref{fig3}(c)]. This higher-order process involves stimulated events which are illustrated by  the vertical arrows in the left panel of Fig. \ref{fig3}(c)    and they  are facilitated on the boundaries of the first and the third Brillouin zone.
 Thus,  we obtain an realization of Umklapp processes which result in the inelastic scattering of electron back into the regime of normal Doppler effect.\section{\label{sec:level2a} Electron bunches and induced Doppler effects}
 \begin{figure}[H]
 \includegraphics[scale=0.6]{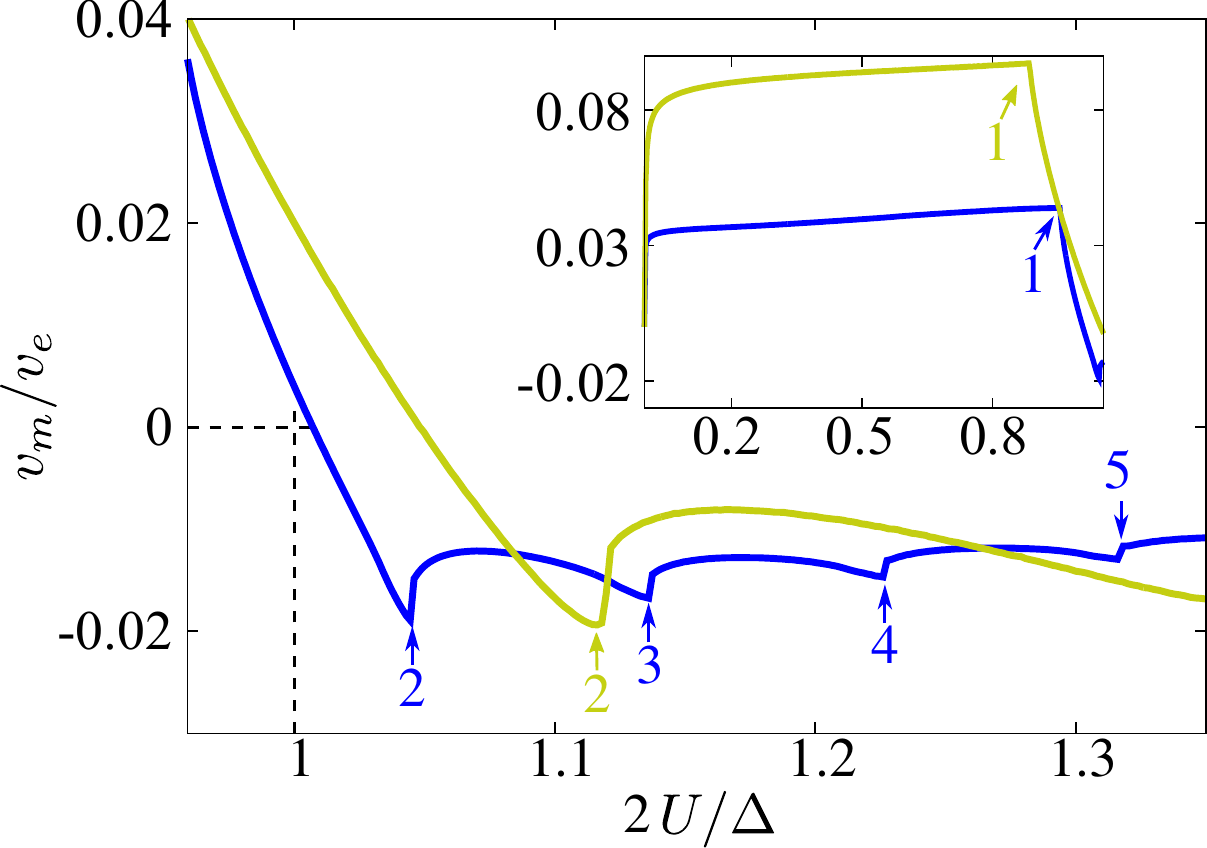}
 \caption{  The  velocity $v_m$ of the electron bunches as a function of the wave amplitude $U$ in the quasi-ballistic regime for $v_e/v_s$=22 (blue curve) and $v_e/v_s$=10.5 (yellow curve); See Table  \ref{tab:secondtable}. The dashed line at $U_0=\Delta/2$ separates the forward motion of the electron bunches from the backward one with respect to the propagation of acoustic wave  for $v_e/v_s\rightarrow\infty$. The comparison with the superluminal selection rules [Eq. (\ref{eq:selectionrules})] is indicated by the arrows. The inset shows the full development of $v_m$ with an increase of $U$ around the first selection rule (anomalous Doppler condition). The acoustic wave in both cases has a frequency lying at $\omega_s=v_s/d$ with $d$ the lattice period of each SL structure.
  } 
 \label{fig4}
 \end{figure}
In this Section we discuss the role of the induced Doppler effects in nonlinear mechanisms, similar to laser-plasma interactions \cite{Rukhadze87, Nezlin76} and relativistic microwave devices \cite{grigoriev2018microwave}, leading to formation of traveling electron bunches.
In the presence of a small-amplitude acoustic wave some of the electrons are captured by the propagating potential, forming a group of trapped electrons (electron bunch)  and perform oscillatory motion within potential itself.  However, if the acoustic stimuli are strong enough they can enforce the electrons to perform complex Bloch oscillations (phonon-assisted trajectories) \cite{greenaway2010using,apostolakis2017nonlinear}, which, as we found out, causes  the electron bunches to oscillate with large amplitude in $p_x$--space.
To understand the role of these effects in the acoustically driven charge transport in SL we analyze the solution of the Boltzmann transport equation (BTE)
\begin{equation}
\dfrac{\partial f}{\partial t}+ v(p_x)\frac{\partial f}{\partial x}+ eE(x,t)\dfrac{\partial f}{\partial  p} = \textrm{St}[f],
\label{eq:bte0}
\end{equation}
 where $\textrm{St}[f]$ is the collision term specified later, $f(x,p_x,t)$ is the electron distribution function, $E(x,t)=E_s \cos(k_s x-\omega_s t)$   is a force field derived from the potential energy function $V(x,t)$  with $E_s=k_s U/e$ denoting the  amplitude of the effective acoustoelectric field.

 
 
We first consider the quasi-ballistic regime $a=k_s l \gg 1$, where $l=v_0 \tau$. Here  $l$ denotes the mean free path of the carriers being sufficiently larger than the sound wavelength and   $\tau$ is the electron  scattering time. In this case, the collision term in (\ref{eq:bte0}) can be neglected as $\textrm{St}[f]\rightarrow 0$.  
 In addition, the value of the sound frequency considered in our calculations  is, $\omega_s=4\times 10^{11}$ rad/s with $k_s\sim 1/d$, which lies far away from the frequency ranges of the phonon stop bands \cite{narayanamurti1979selective, tamura1988acoustic}.
To characterize the kinetic properties  the electrons bunches, we introduce the average momentum $p_a=p_{\phi} \hbar/d$, where 
$p_{\phi}=\textrm{arg}(m_1)$ is a circular mean angle. Here $m_1=\langle \textrm{exp}(ip_x(t) d/\hbar)\rangle_{x_i}$ is the first trigonometric moment $m_1$ of the distribution function \cite{isohatala2012devil,mardia2009directional}. The operator $\langle .\rangle_{x_i}$  performs averaging over an ensemble of time-dependent electron trajectories starting with the same initial momentum $p_i=0$ (low temperature limit)  and different initial positions $x_i\in[-\lambda/2,\lambda/2)$, where $\lambda=2\pi/k_s$ means the acoustic wave length.
\\Figure \ref{fig4}  presents the dependencies of the time-averaged electron velocity $v_m=\langle v_0 \sin(p_a(t)d/\hbar)\rangle_{\Delta t}$  on the wave amplitude $U$ which are numerically calculated  (see Appendix B for more details) for two different ratios $v_e/v_s$. {\color{black} Both graphs demonstrate series of local minima of $v_m$, which are reached at the critical values of $U=U_{cr_n}$. By rearranging terms,  Eq. (\ref{eq:phys-bifur}) can be rewritten in the form
 \begin{equation}
U_{cr_n}=\frac{\Delta}{2}\left[1+(2n-3)\frac{v_s}{v_e}\right],
\label{eq:selectionrules}
\end{equation}
which, as it was mentioned above, in the limit $v_e/v_s\rightarrow\infty$ reveals  a  localization condition 
\begin{equation}
U_{cr_n}\longrightarrow U_0=\frac{\Delta}{2}.
\label{eq:localcon}
\end{equation} These values are highlighted by arrows with the numbers indicating the index $n$ in (\ref{eq:selectionrules}), while the dashed line at $U_0=\Delta/2$ separates the forward motion of the electron bunches from the backward one with respect to the propagation of acoustic wave  for $v_e/v_s\rightarrow\infty$. The graphs evidences that for larger $v_e/v_s$, the $U$-value corresponding to reversing of the electron velocity $v_e$ comes closer to $U_0$, thus confirming the localization criterion (\ref{eq:localcon}). 
\begin{figure}[t]
\centering
 \includegraphics[scale=0.5]{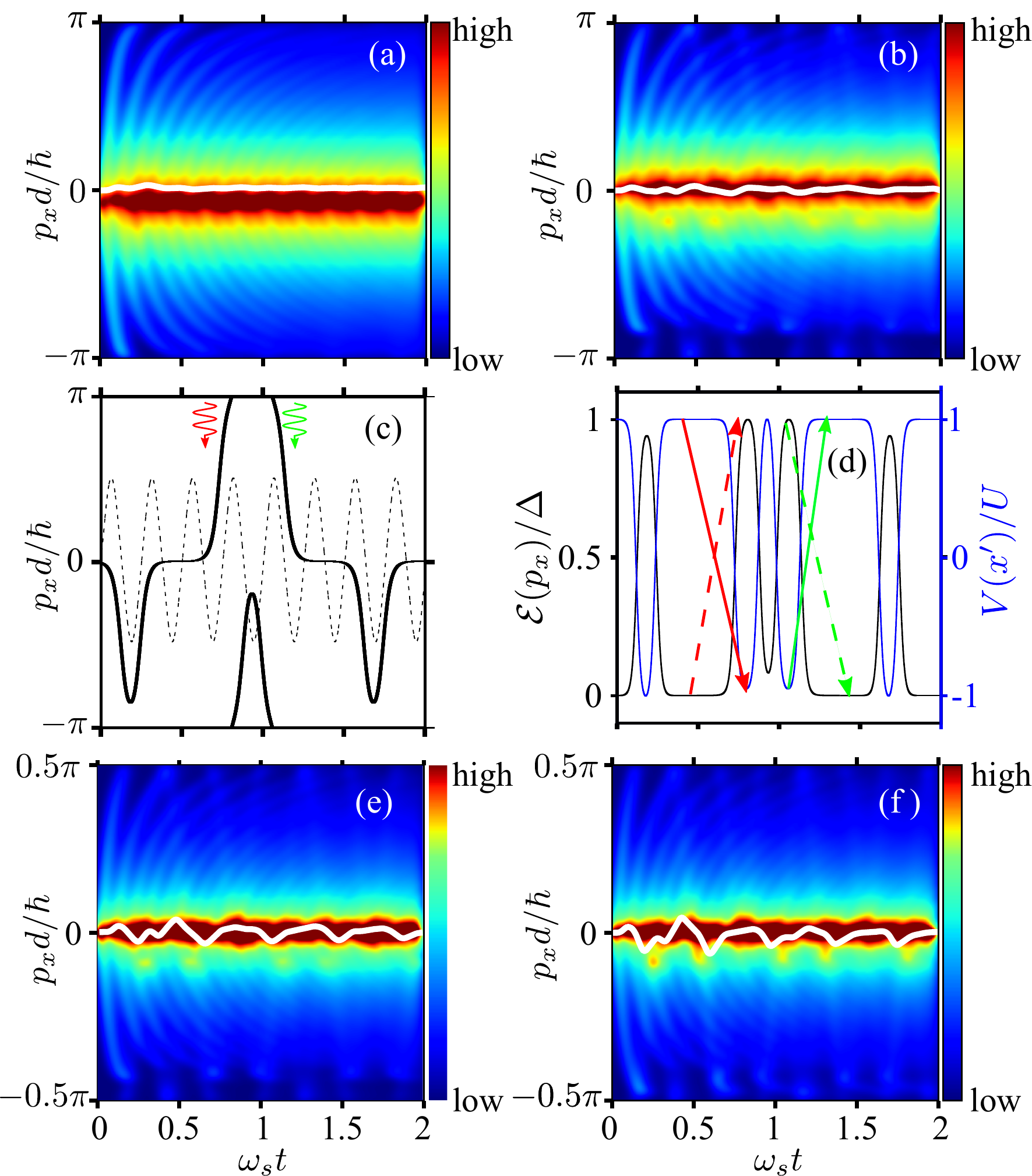}
 \caption{ 
The temporal dynamics of the reduced electron distribution function $\tilde{f}(p_x,t)$ before the first bifurcation (a) $2 U/\Delta=0.95$ and just after (b) $2 U/\Delta=0.98$  with the red region in the colormap  indicating the formation  of the electron bunch.
(c) $p_x$-space trajectories for $2 U/\Delta=0.98$ ($U_{cr_1}<U<U_0$)  and different initial conditions: $x_0=0$, trapped trajectory (dashed line) in the acoustic wave potential  and $x_0 \approx -\pi/(2 k_S)$, trajectory (solid line) which experiences anomalous Doppler instabilities. The wavy arrows indicate the absorption (red) and the emission (green) of phonons. (d) The temporal oscillations of miniband energy (black curve) and potential energy (blue curve)  for $2 U/\Delta=0.98$ ($U_{cr_1}<U<U_0$) and the initial condition $x_0 \approx -\pi/(2 k_S)$. (e), (f) The temporal dynamics of the reduced electron distribution function $\tilde{f}(p_x,t)$ indicating the electron bunch (red) for different wave amplitudes: (c) $2 U/\Delta=1$  ($U=U_0$)  and (d) $2 U/\Delta=$1.04 ($U_{cr_1}<U<U_{cr_2}$). The   white lines designate the  
 mean of electron  quasi-momentum distribution (mean angle $p_{\phi}=p_a d/\hbar$) as a function of time. The calculations have been performed using the parameters of the SL structure with $v_e/v_s=22$.
  } 
 \label{fig5}
 \end{figure} 
\begin{figure}[t]
\centering
 \includegraphics[scale=0.55]{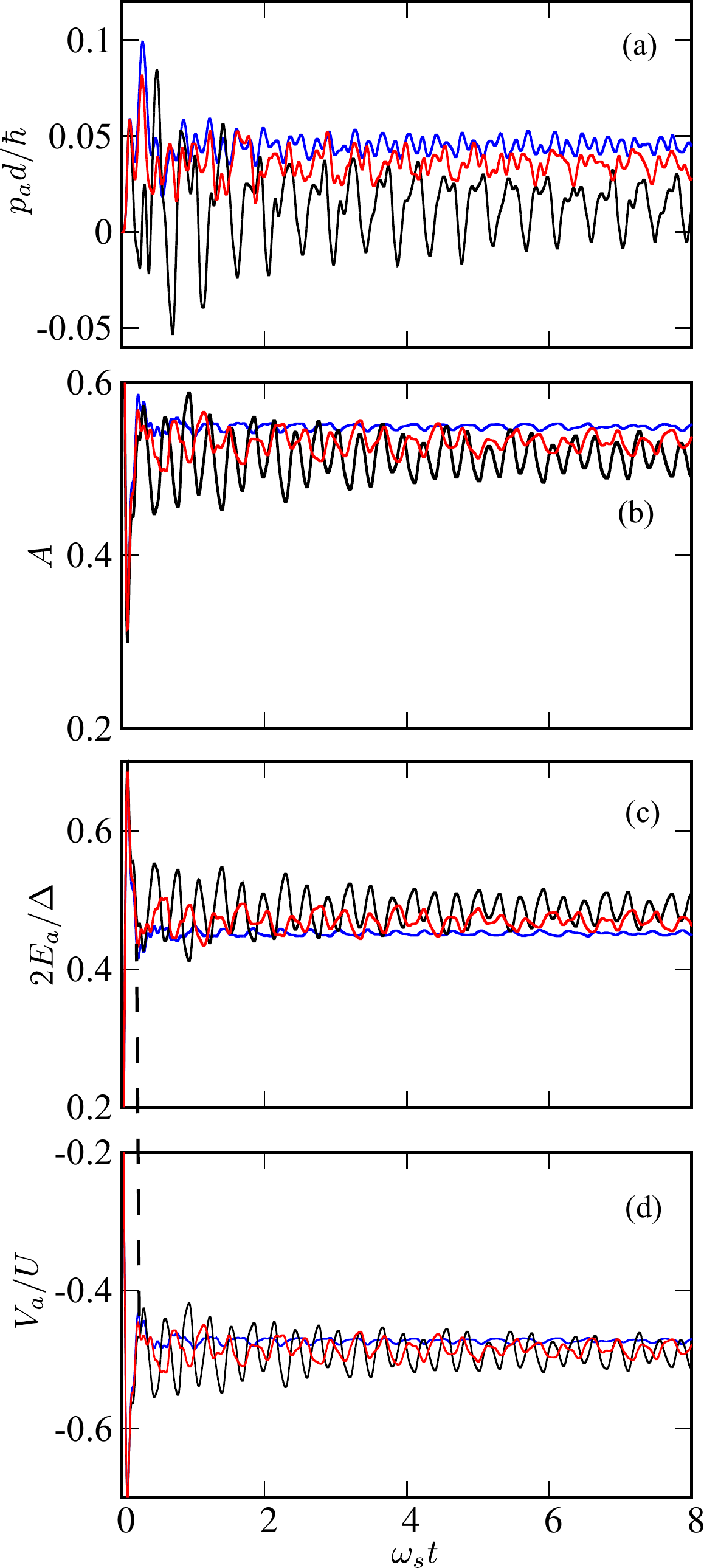}
 \caption{The numerically obtained  oscillations of (a) the averaged momentum $p_a$, (b)  the coherence, $A$, of the electron distribution, (c) the averaged electron energy $E_a$ and (d) the averaged potential energy $V_a$  for $2 U/\Delta=0.95$ ($U<U_{cr_1}$, blue curves), $2 U/\Delta=$0.96, 0.98 ($U_{cr_1}<U<\Delta/2$, black and red curves). The calculations have been performed using the parameters of the SL structure with $v_e/v_s=22$. Dashed vertical line is discussed in text. \color{black}
  } 
 \label{fig6}
 \end{figure} 
\\The Doppler effects as described in the previous section are elementary processes whereas here  the induced Doppler effects correspond to collective processes  triggered by an instability in which an elementary radiation act can induce sequentially another one. This becomes more evident, as we shall see, by the behavior of the electron dynamics between the bifurcations. Figure  \ref{fig5}(a) shows the temporal dynamics of the electron distribution function by solving Eq. (\ref{eq:bte0})  for $U<U_{cr_1}$, while \ref{fig5}(b)  for the case $U_{cr_1}<U<U_0$.   The color in (a), (b) relates to the values of   reduced electron distribution function $\tilde{f}(p_x,t)=1/\lambda\int_{-\lambda/2}^{\lambda/2}f(x,p_x,t)dx$, where it changes from black for the lowest value to red -- for the highest one which indicates the presence of an enhanced concentration  of electron trajectories (electron bunches) around the center of Brillouin zone.  As $U$ increases beyond $U_{cr_1}$, the electron dynamics become more complicated which is  manifested in appearance of quite sophisticated patterns in $\tilde{f}(p_x,t)$, compare Fig. \ref{fig5}(a) for the case of  $U<U_{cr_1}$ and Fig. \ref{fig5}(b) for the case of  $U>U_{cr_1}$. While the former figure demonstrate rather regular pattern, the latter shows signatures of turbulent behavior, which evidences e.g in occurring isolated color spots on the map. Such dramatic change in electron distribution can be attributed to emergence of a specific phonon-assisted  trajectories (complex Bloch oscillation) which is shown as a solid curve in Fig. \ref{fig5}(c), recounting a wave packet moving slowly in the proximity of $p_x=0$ and quickly at $p_x=\hbar\pi/d$. This type of trajectories appear for $U>U_{cr_1}$ in addition to another low-amplitude regular $p_x$-trajectory [dashed curve, Fig. \ref{fig5}(c)] which existed also for $U<U_{cr_1}$.  The latter one (dashed curve) indicates a wave packet moving quickly around the center of the Brillouin zone, corresponding to the confined electron motion within the propagating moving potential. These two types of trajectories differ by their initial values $x_i$. The motion of the electrons starting from certain initial conditions such as the solid trajectory in Fig. \ref{fig5}(c) experiences  Doppler instability accompanied either by phonon absorption (red wavy arrow) or emission (green wavy arrow) which give rise to oscillations of electron bunches with larger amplitude.  Figure \ref{fig5}(d) illustrates the balance between kinetic $\mathcal{E}$ and potential energy $V$ for phonon absorption or emission processes. In an absorption process, a wave packet with a small crystal momentum $p_x \sim 0$, gains a considerable portion of energy  while moving [red dashed arrow, Fig. \ref{fig5}(d)] from the bottom to the top of the miniband  at the expense of potential energy  [red curve, Fig. \ref{fig5}(d)]. In contrast, during the emission process, a wave packet with a large $p_x$ (near $p_x=\pi\hbar/d$) loses a considerable part of its energy [green dashed arrow, Fig. \ref{fig5}(d)] while a making a transition from the top to the bottom of the potential (green arrow, Fig. \ref{fig5}(d)). Each radiation act is preceded (followed) by extended presence of electron   around the center of the Brillouin zone.  Considering now an electron ensemble, the electron wave packets are concentrated around a single trajectory [white curve in Fig. \ref{fig5}(b)] with mean value $p_a$ that  oscillates further close to the center of BZ in comparison to the trajectory in Fig. \ref{fig5}(a), causing the electron bunches to slow down; See the sudden drop of $v_m$  in the inset of Fig. \ref{fig4}.  At $U=U_0$ the electrons eventually become localized. In this case, the averaged momentum $p_a$ [white curve in Fig. \ref{fig5}(e)]  oscillates almost periodically around $p_x$=0 resulting in  a zero $v_m$. For $U>U_0$ the electronic bunches counter-propagate with respect to the  propagating sound wave.   Such inversion of motion  becomes apparent in Fig. \ref{fig5}(f), where the electron bunch is shifted below $p_x=0$. The inversion of electrons drift} can be explained by an increase of electron trajectories 
  which are subjected to anomalous Doppler  shifts. The backward drift of electrons becomes maximal at $U_{cr_2}$, coinciding with the first manifestation of normal Doppler effect as discussed in Sec. \ref{sec:level2}. \\To gain a deeper insight in the   emission processes and their implications in electron bunching, we examine in detail  the dynamics of averaged parameters  in the vicinity of the first bifurcation at $U\approx U_{cr_1}$. Namely, we analyze the characteristics, which beside the averaged momentum $p_a$ (Fig. \ref{fig5}) include the coherence of electron distribution $A=|m_1|$, the averaged electron energy, $E_a = \Delta/2 \langle 1-\cos\lbrace p_x(t)d/\hbar\rbrace \rangle_{x_i}$, and the averaged potential energy $V_a=-U\langle \sin(k_s x-\omega_s t) \rangle_{x_i}$.  Note that  the electron distribution converges to a Dirac distribution centered  on a single trajectory at mean value $p_a$ for  $A=0$  whereas for the opposite limit $A=1$  there is no well defined mean of the electron momentum. Figure \ref{fig6}(a) shows the dynamics of $p_a$ in time as $U$ growth. For $U<U_{cr_1}$= 9.55 meV ($2 U/\Delta=0.955$), $p_a$ demonstrates a low-amplitude erratic fluctuations (blue curve), which are substituted by more regular close to quasi-periodic oscillations once $U$ exceeds  $U_{cr_1}$ (red curve). Further growth of $U$  leads to increase of the amplitude of the oscillations that remains quite regular (black curve). Such regularization of the electron bunch dynamics and appearance of pronounced large-amplitude oscillations of $p_a$ is associated with the Doppler instability evoked by he absorption or the emission of phonons, which, as it has been shown above, take place for $U>U_{cr_1}$.  Repeating inelastic absorption or emission events perturb also the coherence parameter $A$ which demonstrates larger oscillations for larger $U$, indicating that electron bunches are created and then deformed almost periodically (compare blue, red and black curves in Fig. \ref{fig6}(b)). This further explains the abrupt change in electron distribution and the formation of distinct color spots with time variation as was demonstrated in Fig. \ref{fig5} (e), (f). The effect of the Doppler instability related to phonon exchange can be clearly seen in the time realizations of the kinetic $E_a$ and potential  component  $V_a$ of the electrons energy, which are presented in Fig. \ref{fig6} (c) and (d), respectively. For $U< U_{cr_1}$, the energy the exchange between $E_a$ and $V_a$ is very small (see blue curves in (c) and (d)). However, if $U$ even slightly exceeds $U_{cr_1}$, the interplay between $E_a$ and $V_a$ dramatically intensify, and become larger for larger $U$ (see red and black curves in (c) and (d)). The vertical dashed lines in (c) and (d) illustrates the energy transformation. Namely, it indicates a moment when the electron bunch having a large potential energy $V_a$ start to lose it due to phonon exchange events with simultaneous increasing the kinetic component $E_a$. We note that the concept of a wave-like  bunching of the electrons \cite{kroemer2000nature,schomburg2003amplification,demarina2005bloch} in momentum space has already  been used to explain the amplification of a THz field which might arise due to the interaction of the bunches with the THz field itself. Extended  analysis of the effects of the scattering processes on miniband transport and sound absorption will be presented in the following section.
 
\section{\label{sec:level3} NONLINEAR ELECTRIC TRANSPORT AND THE SOUND ATTENUATION EFFECTS }

\begin{figure}
\includegraphics[scale=0.35]{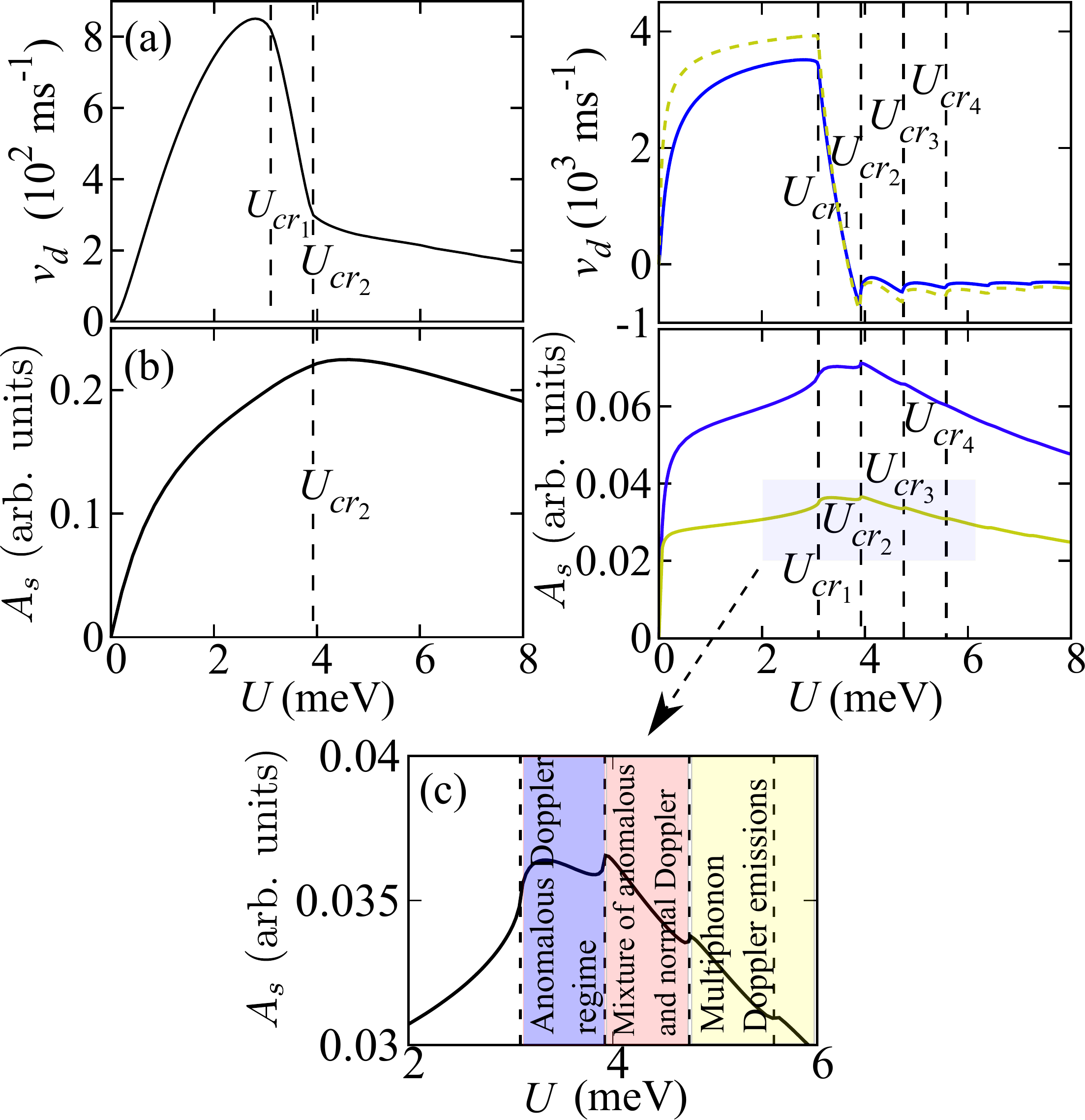}
   \caption{(a) The dependence of the drift velocity $v_d$ as a function of the acoustic wave amplitude $U$. The left panel demonstrates $v_d(U)$ for the wave frequency $\omega_s=0.1\tau^{-1}$  whereas  the right panel for $\omega=0.5\tau^{-1}$ (dashed yellow curve) and $\omega_s=\tau^{-1}$ (blue curve). (b) The dependence of the absorption coefficient of the sound wave $A_s=\Gamma/(2k_sU)$ upon $U$ calculated for different frequencies: $\omega_s=0.1\tau^{-1}$ (left panel), $\omega_s=0.5\tau^{-1}$ (right panel, blue curve) and $\omega_s=\tau^{-1}$ (right panel, yellow curve). (c) Zoom in of (b)--right panel, showing the different regimes of phonons emissions.  The vertical lines designate the critical values of $U$. The calculations have been performed using the parameters of the SL structure with $v_0/v_s=13$; See Tables \ref{tab:secondtable}, \ref{tab:thirdtable}. }
\label{fig7}
\end{figure}
Up to now, we have considered the quasiballistic regime, i.e. $\alpha=k_s l\gg1$, to study the role of the acoustoelectric effects in the quasi-classical description of the electron transport. However, the effects of coherent phonon dynamics on the miniband electrons and the related superluminal phenomena might appear over a wide range of the parameter $\alpha$,  as it was previously shown for electron phonon interactions  in metals \cite{spector1967interaction,abrikosov2017fundamentals}.  A more general approach is to take into account the scattering effects, when the mean free path, $l=v_0 \tau$, is comparable with to the sound wavelength $\lambda$. In this case, to describe the directed electron transport in a superlattice we  use the Boltzmann transport equation  (\ref{eq:bte0}) 
with  $\textrm{St}[f]=(f_0-f)/\tau$, where $\tau$ is the relaxation time and $f_0(x,p_x)$ is the Fermi distribution in the nondegenerate limit. Solution of BTE yields the drift velocity of electrons
\begin{equation} 
 v_d=\int_0^{T_s}\frac{dt}{T_s}\int_{-\infty}^t e^{\frac{-(t-t_i)}{\tau}}v_x(t,t_i)\frac{dt_i}{\tau} \label{eq:Boltz_vd}. 
\end{equation}
$t_i$ is the time of last collision at position $x_i$, $v_x(t,t_i)$ is a trajectory in time governed by Eqs. (\ref{eq:all-dot}) and $T_s$ $=2\pi/\omega_s$ is the period of the acoustic plane wave. The relaxation rate approximation, $\tau$, can facilitate both elastic scattering processes \cite{fromhold2004chaotic} and additional inelastic scattering  processes due to long wave phonons. Eq. (\ref{eq:Boltz_vd})
 can be obtained using the time-dependent path-integral approach \cite{Budd63-PI,maccallum1963kinetic,bass1981theory} which allows to investigate generic features of electron transport and unravel the acoustolectirc effects that might arise in the system. To analyze the deformation propagation in a SL we additionally take into account the back action of electrons on the phonon wave by considering the sound attenuation effects. In general, the electron contribution to the attenuation of the ultrasound in materials arises because  energy is transferred between the wave and electrons.    Therefore, the absorption coefficient of the sound is given by \cite{Kazarinov63,blount1959ultrasonic,gal1979nonlinear}
\begin{equation} 
\Gamma= \int \left<\dot{H}(p_x,t)f(p_x,x,t)\right>_{T_s} dp_x.
\label{eq:pabs}
\end{equation}
Here  $\dot{H}=dH/dt$, whereas the angle brackets designate averaging over time, namely the period of the sound wave $T_s=2\pi/\omega_s$. By solving Eq. (\ref{eq:pabs}) we find (see Appendix \ref{App1}) the absorption coefficient of the sound in the SL miniband, which is presented for our convenience in a normalized form
\begin{equation} 
A_s=\left<\left(\tilde{v}(t)-\dfrac{v_s}{2v_0}\right)\cos(k_s x-\omega_s t)\right >_{T_s},
\label{eq:acoe}
\end{equation}
where the absorption coefficient $\Gamma$ is related to $A_s$  as  $\Gamma=(2 k_s U)A_s$ and $\tilde{v}(t)=\left<{v}_x(t)\right>/v_0$ 
is the electron velocity averaged over the distribution function, $f$, satisfying 
 the BTE [Eq. (\ref{eq:bte0})]. The dependence of the drift velocity $v_d$ on $U$ for an acoustic wave with $\omega_s=0.1\tau^{-1}$ $(\alpha\sim1.3)$ and a superlattice with a miniband width of 7 meV ($v_0/v_s=13$) is presented in the left-hand panel of Fig. \ref{fig7}(a). One can recognize two characteristic values of $U$. First, the drift velocity is drastically suppressed beyond $U=U_{cr_1}$ and subsequently the $v_d(U)$ characteristic exhibits an observable change in slope at $U=U_{cr_2}$.
The black solid curve in Fig.  \ref{fig7}(a) reminds the classical Esaki-Tsu $v_d(E_{dc})$  \cite{esaki1970superlattice}, which describes the response of miniband electrons to an electric field, $E_{dc}$, applied along the growth direction  of the superlattice. However, in our case the specific changes in $v_d$ with variation of $U$ are directly associated with changes in the electron bunches dynamics, revealing  the appearance of the superluminal effects as discussed in Sec. \ref{sec:level2a}. Hence, the critical amplitudes $U_{cr_1}$ and $U_{cr_2}$ designated by the vertical lines in Fig. \ref{fig7}(a) correspond to the thresholds for triggering the superluminal anomalous and superluminal backward Doppler effects that are associated with emission and absorption of phonons  in the presence of scattering. Figure  \ref{fig7}(b)  demonstrates the behavior of $A_s$ with change of $U$. The increase of $U$ first enhances the absorption of the acoustic wave, since its its interaction with the electrons becomes stronger. However, once $U$ reaches $U_{cr_2}$ the absorption starts to drop gradually. Such reduction of $A_s$  relates to  the rise of normal Doppler effects resulting from the recoil momentum given to the radiating system by the emitted phonons.  The changes in $v_d(U)$  and $A_s(U)$ indicating  transitions between different superluminal regimes which become more evident  for acoustic driving with higher frequency $\omega_s$. In the right panel of Fig. \ref{fig7}(a) the dependencies $v_d(U)$ are shown for  $\omega_s=0.5\tau^{-1}$ $(\alpha\sim 6.6)$  by dashed yellow curve and $\omega_s=\tau^{-1}$ $(\alpha\sim13.2)$ by solid blue curve. Both dependencies have pronounced features  at the same  values of $U$. However, in contrast to the low-frequency case presented in the left panel Fig. \ref{fig7}(a) the high-frequency dependencies indicate more features also related to $U_{cr_n}$  (\ref{eq:u_cr}) for $n>2$. Importantly, these high-frequency (short-wavelength) acoustic excitations can induce the reverse of the electron drift ($v_d<0$) due to mechanisms discussed in Section \ref{sec:level2}.
The features in the graph of $v_d(U)$ are also reflected in the dependence $A_s(U)$ shown in in the right-hand panel of Fig. \ref{fig7}(b). For example, a kink in  $A_s(U)$ at $U_{cr_1}$ in Fig. \ref{fig7}(b) coincides  with the maximum of $v_d(U)$  in Fig. \ref{fig7}(a). A local minimum of $A_s$ at $U=U_{cr_2}$ corresponds to the reverse of drift velocity $v_d$. Other abrupt changes in $v_d$ are observed when $A_s(U)$ attains the  minima [see  Figure  \ref{fig7}(b)]  at $U_{cr_n}$ for n=3 and 4.  The appearance of these kinks, i.e. local minima in $v_d(U)$, $A_s(U)$ can attributed to the transition between different supeluminal regimes of phonons emission.
 Figure \ref{fig7}(c) illustrates these regimes considering a zoomed part of yellow curve $A_s(U)$ from the right panel of the Fig. \ref{fig7}(b). They include:
  (i) an anomalous Doppler regime ($U_{cr_1}<U<U_{cr_2}$) indicated by the blue shaded area in Fig. \ref{fig7}(c) where a significant portion of electron trajectories experiences anomalous Doppler instabilities which start to move backwards leading to the suppression of drift velocity and considerable reduction in the sound absorption coefficient.  (ii) Coexistence of anomalous and normal  Doppler effects ($U_{cr_2}<U<U_{cr_3}$). In this regime, an increase of $U$ enables more charged particles to experience  normal Doppler instabilities resulting in emission of phonons at the expense of electron  trajectories which are subjected only to anomalous Doppler instabilities. Hereafter, the overall drift of electron bunches remains negative   whereas the absorption of the sound wave is further reduced. (iii) multiphonon Doppler processes ($U>U_{cr_3}$) in which the electrons are inelastically scattered by the acoustic wave allowing  their trajectories to enter successively higher Brillouin zones.  These more complicated trajectories are responsible for the further suppression of the absorption coefficient.  

\section{\label{sec:level4} Absolute negative mobility}

\begin{figure}[t]
 \includegraphics[scale=0.6]{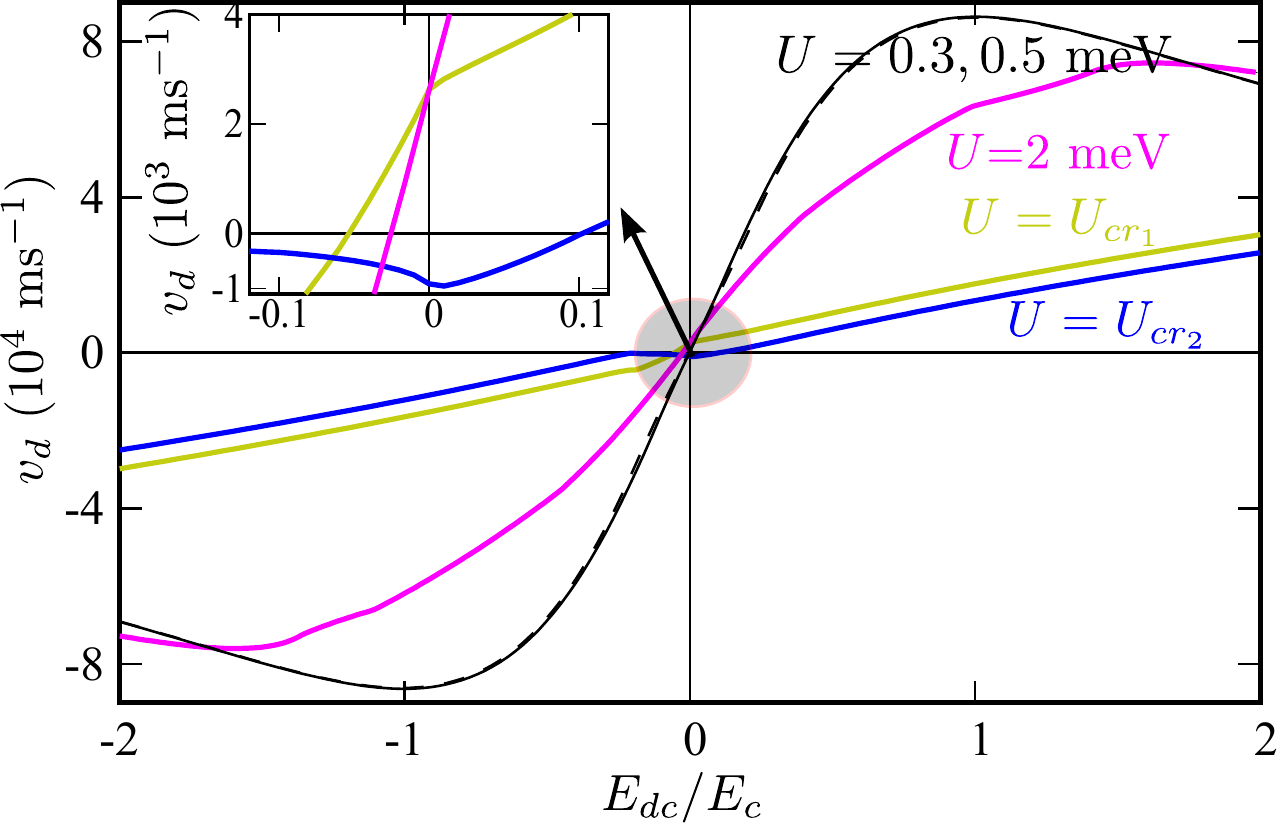}
 \caption{ Drift velocity-electric field characteristics with variation of the wave amplitude $U=0.3, 0.5, 2 $ meV and at the critical values of the instabilities $U_{cr_1}=9.55$ meV and  $U_{cr_2}=10.46$ meV. $E_c=2.3$ kV/cm is the critical field for the onset of NDV in the Esaki-Tsu characteristic. The solid and dashed black lines indicate the $v_d(E_{dc})$ characteristics for small wave amplitude at $U=0.3$ meV and $U=0.5$ meV respectively. The frequency of the acoustic wave is fixed at $\omega_s=0.1 \tau^{-1}$ ($\alpha$ $\sim$ 3.5) The calculations have been performed using the parameters of the SSL structure with $v_0/v_s=35$; See Table \ref{tab:secondtable}.
  } 
 \label{fig8}
 \end{figure}

\begin{figure}[t]
 \includegraphics[scale=0.6]{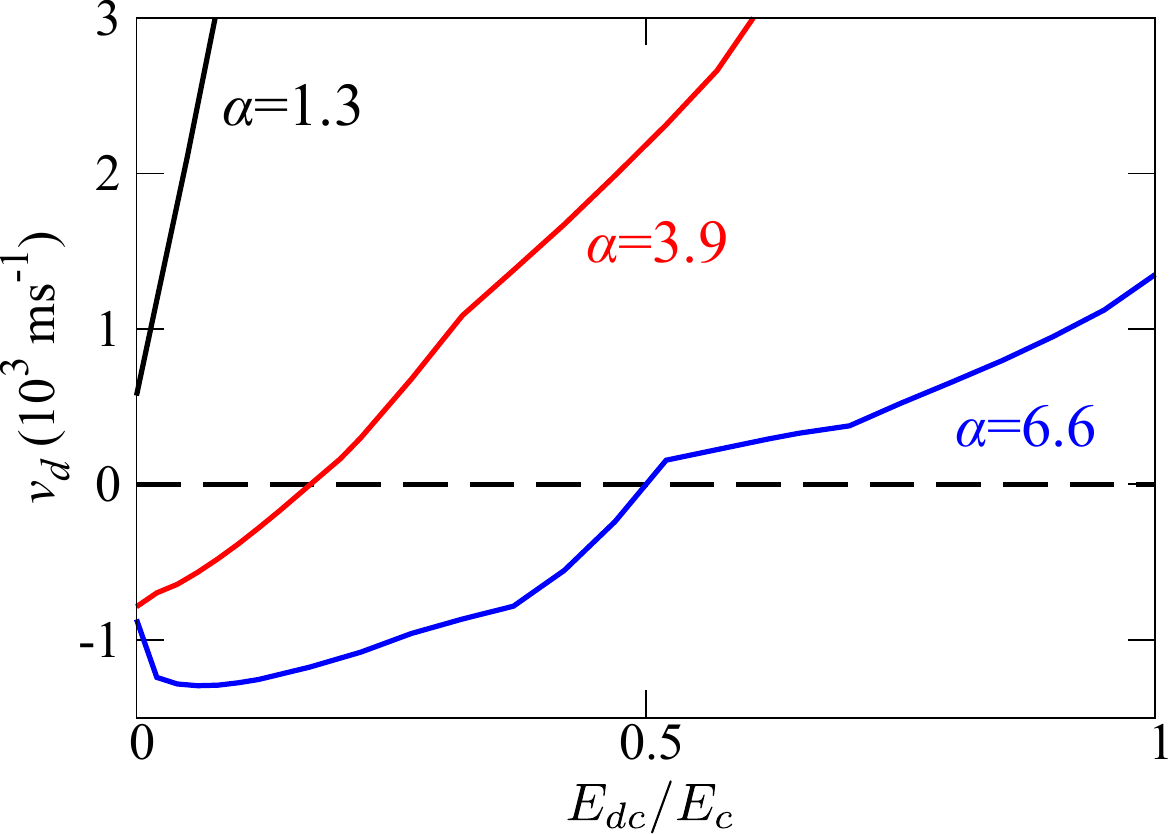}
 \caption{ Dependence of the drift velocity $v_{dc}$ on the dc bias $E_{dc}$ for the amplitude of the acoustic wave $U=U_{cr_2}=3.92$ meV and for different sound frequencies  : positive slope for $\omega_s\tau=0.1$ ($a=1.3$, black curve), ANM for   $\omega_s\tau=0.3$ ($a=3.9$, red curve) and   $\omega_s\tau=0.5$  ($a=6.6$, blue curve). $E_c=2.1$ kV/cm is the critical field for the onset of NDV in the Esaki-Tsu characteristic. The calculations have been performed using the parameters of the SSL structure with $v_0/v_s=13$; See Table \ref{tab:secondtable}.
  } 
 \label{fig9}
 \end{figure}

 In previous Section  \ref{sec:level3}, we studied the superluminal mechanisms governing the directed electron transport in a strongly coupled SL subjected to a sole acoustic plane wave. Next,  we consider the  absolute negative mobility (ANM)  and the superluminal Doppler effects which go hand in hand in case  a constant electric field $E_{dc}$ is additionally applied along the SL.  The ANM has been already reported \cite{keay1995dynamic} in  DC-biased SLs under high frequency  irradiation $\propto$ $E_{\Omega}\cos(\Omega t)$, shown as  inversion of electron current in the vicinity of $E_{dc}=0$. In our model, the acoustic drive oscillates both in time and space and thus the  semiclassical equations of motion for the miniband electrons are
\begin{subequations}
\label{eq:all-dot2}
\begin{eqnarray}
v&=&v_0\sin\left(\frac{p_xd}{\hbar}\right),  \label{xdot2}\\
\frac{dp_x}{dt}&=&eE_{dc}+k_s U\cos[(k_s(x+x_i) -\omega_s t)], \label{pdot2}
\end{eqnarray}
\end{subequations}
where $e>0$ is the elementary charge. 
Figure  \ref{fig8} shows the dependencies of drift velocity $v_d$ as a function of the electric field $E_{dc}$ numerically calculated for different amplitudes of $U$. For our calculations we chosen the parameters close to ones from recent experiments \cite{shinokita2016strong},  $\Delta=20$ meV and  $\omega_s=0.1\tau^{-1}$ ($\alpha$ $\sim$ 3.5). For a weak acoustic excitation (black solid and black dashed curves) $v_d(E_{dc})$ is almost identical to the Esaki-Tsu dependence \cite{esaki1970superlattice}.  Therefore, the supersonic condition ($v_d>v_s$) for the Cherenkov emission can exist for $U \ll U_{cr_1}$, however it does not necessary produce ANM or result in the suppression of the drift velocity. This was confirmed in the recent experiments  \cite{shinokita2016strong}, which have shown that the current-voltage characteristics are almost identical with or without illumination of the SL sample by the femtosecond pulse train resulting the coherent acoustic wave generation  due to stimulated Cherenkov phonon emission. However, as the value of $U$ increases the dependence become less and less steeper, compare pink curve for $U$= 2 meV, yellow curve $U=U_{cr_1}$=9.55 meV and  blue curve for  $U=U_{cr_2}$=10.46 meV. The mechanisms of such a suppression  \cite{poyser2015weakly,wang2020ultrafast} are similar to the mechanisms governing the electron transport in a DC-biased SL under electromagnetic  irradiation \cite{keay1995dynamic}. They relate to the fact that the electron tunnelling probability is affected by the energy $\hbar \omega$ of the excitation quantum of frequency $\omega$, which could be either photon ($\hbar\Omega$) or phonon ($\hbar\omega_s$) interacting with the electron. Here, the drift velocity can be expressed in terms of the phonon-assisted replicas of the Esaki-Tsu drift velocity \cite{Wacker20021,poyser2015weakly},
\begin{equation}
v_d \sim \sum_{n=-\infty}^{\infty} J_{n}^2(\beta)v_{d_0}(E_{dc}d+n\hbar \omega_s/e), 
\label{eq:anal_vd}
\end{equation}
where  $J_n$ is the first-kind Bessel function of order $n$, $v_{d_0}(E_{dc})$ is the dependence of the electron drift velocity on the DC bias $E_{dc}$ for SL in the absence of  photon/phonon excitations, $\beta=edE_{s}/(\hbar\omega_s)$ and $E_s=$ $k_s U/e$ is voltage equivalent of the magnitude of the acoustic wave driving the SL. For a range of $\beta$ the term with  $J_0$ produces a dominant effect, so implying suppression of the $v_d$ with growth of $U$. The validity of Eq. (\ref{eq:anal_vd}) for the phonon-assisted transport is limited either to small amplitude $\beta$  (black curve, Fig. \ref{fig8})
or to the quasistatic limit $a=k_s l \ll 1$ \cite{lax1965microwave}. Thus, we resort instead to the fully numerical integration
of the model (\ref{eq:all-dot2}) and calculation of drift velocity via BTE solution (\ref{eq:Boltz_vd}) which indicate that ANM takes also place for a propagating acoustic excitation.  In particular, we found that for selected values of the SL parameters, ANM  is realized when $U$ exceeds $U_0=10$ meV. Since $U_{cr_1}<U_0<U_{cr_2}$, it indicates the gradual onset of the ANM due to the absorption and emission of phonons  under the conditions of the anomalous Doppler effect. The  inset of Fig.  \ref{fig8} illustrates a zoomed part of the main panel of Fig.  \ref{fig8} in vicinity of $E_{dc}=0$ and it reveals a negative drift of the electrons ($v_d<0$) for $E_{dc}=0$ when $U=U_{cr_2}$=10.46 meV (blue curve) and positive $v_d$ when $U=U_{cr_1}$=9.55 meV  (yellow curve) or $U=$2 meV (pink curve). It follows that the condition for near zero mobility of an acoustically driven SL  can be given by 
\begin{equation}
E_s=\alpha E_c.
\label{eq:ANMcon}
\end{equation}
Here, we have rewritten localization condition (\ref{eq:localcon}) in terms of the amplitude, $E_s$ of the effective acoustoelectric field, whereas $E_c=\hbar/(ed\tau)$ is the critical electric field for the onset of the negative differential velocity (NDV) in Esaki-Tsu characteristic \cite{esaki1970superlattice}. In contrast, the localization condition \cite{keay1995dynamic, Wacker20021} for the photon-assisted transport is $J_0(\beta)=0$ with the first root given at $\beta=E_{\Omega}/(E_c\Omega\tau)\sim 2.4$. Furthermore, the dynamic localization and ANM are not possible for any ac field amplitude when $\Omega\tau<1$ and therefore the smallest amplitude of an ac field which can induce ANM is $E_{\Omega}=2.4 E_c$ at $\Omega=1/\tau$ \cite{ignatov1995thz}. In Sec. \ref{sec:level3} and Appendix \ref{App5}  we noticed that Doppler effects start to have more prominent implications in directed transport once the sound wavelength becomes comparable with the mean free path of the charge carrier, i.e. $\alpha/(2\pi)\gtrsim0.5$. This  is further confirmed by  Fig. \ref{fig9} showing that the increase of sound frequency $\omega_s$ and therefore of $\alpha$ results in the appearance of ANM ($\alpha=3.9$, red curve) similar to Fig. \ref{fig8} in which we have considered a slightly smaller $\alpha=3.5$. Hereafter, Eq. (\ref{eq:ANMcon})  indicates values that are comparable with the minimum amplitude of ac field  which can induce ANM but with considerably lower oscillating frequencies. On the other hand, for $\alpha$ close to zero, i.e low 
 sound frequencies  $\omega_s$
 or SL structures with extremely narrow minibands \cite{grahn1991electrical} and therefore  smaller $v_0$, onset of ANM is not feasible. In this limit, the acoustic wave driving acts practically like a quasistatic ac signal, whereas the electron drift can be described well by (\ref{eq:anal_vd}).
\section{\label{sec:level5} B\lowercase{roadband amplification of} EM  \lowercase{waves}} 
In this Section we consider amplification of high-frequency electromagnetic waves in acoustically driven SLs [see Fig. \ref{fig1}(b)]  and the corresponding role of the Doppler effects, which result in the gain similar to the Bloch gain in electrically biased SLs \cite{Ktitorov72,ignatov1976nonlinear}. 
In this study, we consider the model (\ref{eq:all-dot2}), in which the Eq. (\ref{pdot2}) is substituted by
\begin{equation}
\frac{dp_x}{dt}=eE_{\omega} \cos(\omega t)+k_s U\cos[(k_s(x+x_i) -\omega_s t)], \label{pdot3}
\end{equation}
with the term $E(t)=E_{\omega} \cos(\omega t)$ designating a weak probe field of the amplitude $E_{\omega}$ and frequency $\omega$. Note that in practice, the frequency $\omega$ can be favorably tuned by an external resonant cavity \cite{renk2005subterahertz}. The absorption of the probe ac field $E(t)$ is determined  by the real part of the dynamical conductivity \cite{Hyart2009}
\begin{equation}
\sigma_r(\omega)=\dfrac{2\langle j(t)\cos(\omega t)\rangle}{E_{\omega}}, \label{rdc}
\end{equation} where $j(t)=eNv_d(t)$ is the time-dependent current generated by the SL driven by the acoustic wave which is calculated using the time-dependent drift velocity
\begin{equation}
 v_d(t)=\int_{-\infty}^t e^{\frac{-(t-t_i)}{\tau}}v_x(t,t_i)\frac{dt_i}{\tau}, \label{tvdc}
 \end{equation} 
 similar to the semiclassical approach presented in Sec. \ref {sec:level3}, whereas the   ballistic trajectory $v_x(t,t_i)$ is now governed  by Eqs. (\ref{xdot}), (\ref{pdot3}).  Within this framework, the absorption corresponds to $\sigma_r(\omega)>0$ and gain to $\sigma_r(\omega)<0$. The Drude conductivity of the SL is $\sigma_0=2j_p/E_{c}$ where $j_p=eNv_0$ is the peak current density,  directly proportionally to electron density $N$.  Here, $\sigma_r(\omega)$ is estimated for a superlattice  with a miniband widh (20 meV) sufficiently smaller than the optic phonon energy  in GaAs \cite{blakemore1982semiconducting} and a pump frequency $\omega_s\tau=0.1$. The magnitude of the  acoustic wave's  frequency corresponds to $\alpha$=3.5. The proper choice of $\alpha$--parameter  is important for the appearance of  electron bunches with a negative drift and ANM,  which indicate strong involvement of Doppler effects,  and for controlling interaction between bunches and the probe field, see Appendix \ref{App5} for more details.
 \\Figure \ref{fig10}  illustrates how the absorption profile $\sigma_r(\omega)$ is affected by the variation of  $U$.  For small wave amplitude $U$=1.2 meV, $\sigma_r(\omega)$ (green dash-dotted curve) almost follows the free-carrier absorption (red curve) in the absence of a pump field,  demonstrating a power-like decay as probe frequency $\omega$ increases, i.e. $\sigma_r \varpropto (1+\omega^2\tau^2)^{-1}$. Comparison of the  absorption profiles calculated for $U=U_{cr_1}$ (yellow), $U=U_0$ (black) and $U=U_{cr_2}$ (blue) shows that as $U$ grows, the low-frequency absorption gradually decreases, and for $U$ close to $U_{cr_2}$ a low-frequency gain is realized in the system, see blue curve for $U=U_{cr_2}$.  This transition originates from the emergence of  phonon-assisted Bloch oscillations which are associated with the anomalous Doppler effect.  As we showed earlier in Section \ref{sec:level2}, at $U \approx U_{cr_1}$ the electrons starts to demonstrate the Doppler instabilities 
that obey the first selection rule [see Eq. (\ref{eq:phys-bifur})].
As a result electron bunches are formed, which include the frequency-modulated Bloch oscillations with maximal frequency $\omega_B^{max}=k_sUd/\hbar$  \cite{apostolakis2017nonlinear,greenaway2010using}. Appearance of such trajectories decreases absorption for low frequencies $\omega$. Indeed, the absorption profile calculated for $U = U_{cr_1}$ (yellow curve) shows significantly lower low-frequency absorption as compared to one for $U=1.2$ meV (green curve), with the second maximum of absorption in the vicinity of $\omega_B^{max}$, which for the given parameters corresponds to $\omega\tau\approx3.3$. Further increase of $U$ can localize electrons drift, see the inset illustrating the averaged ballistic electron trajectories calculated for $U=U_{cr_1}$ (yellow) and for $U_0$ (black). This localization leads to further decrease of absorption for low-frequency range of electromagnetic probe field, compare black and yellow curves in the main panel of Fig. \ref{fig10}.  For large enough $U$, the Doppler effects become more and more prominent, and the drift of electron bunches can eventually be reversed. These mechanisms create a condition for appearance of a low-frequency gain ($\sigma_r(\omega)<0$) as it is evidenced by the absorption profile calculated for $U=U_{cr_2}$ (blue line). The numerical calculations reveal that for the chosen parameters and $U=U_{cr_2}$, the gain is possible for the wide range of frequencies up to $\omega/(2\pi)=$ 480 GHz. Further  analysis showed that $\sigma_r(\omega)$ can obtain a significant THz gain value ($\sigma_r \approx -0.07 \sigma_0$)  at  $U=U_{cr_2}$  comparable with the maximum gain of a merely dc-biased superlattice (i.e. canonical Bloch oscillator): min$\lbrace \sigma_r(\omega)/\sigma_0\rbrace=-1/8$ at $E_{dc}/E{c}=\sqrt{3}$ \cite{Hyart2009} and larger than expected from SLs driven by  monochromatic and polychromatic fields with oscillating frequencies at the sub-THz range \cite{Hyart08}.  Normally, we relate the magnitude of the gain $\alpha_g$  in units c$\textrm{m}^{-1}$ to dynamical conductivity as $a_g=a_0(\sigma_r/\sigma_0)$ with $a_0=1/(c\sqrt{\epsilon} \epsilon_0)\times 2j_p/E_c$ \cite{wacker2002gain,Hyart08}. This means that one can acquire the value $a_g=37$ c$\textrm{m}^{-1}$  for $a_0\approx527$ c$\textrm{m}^{-1}$ given a  temperature of few millikelvins,  moderate doping $N=10^{16}$ c$\textrm{m}^{-3}$ and relative permittivity $\varepsilon=13$ corresponding to GaAs.   
\begin{figure}[t]
   \includegraphics[scale=0.65]{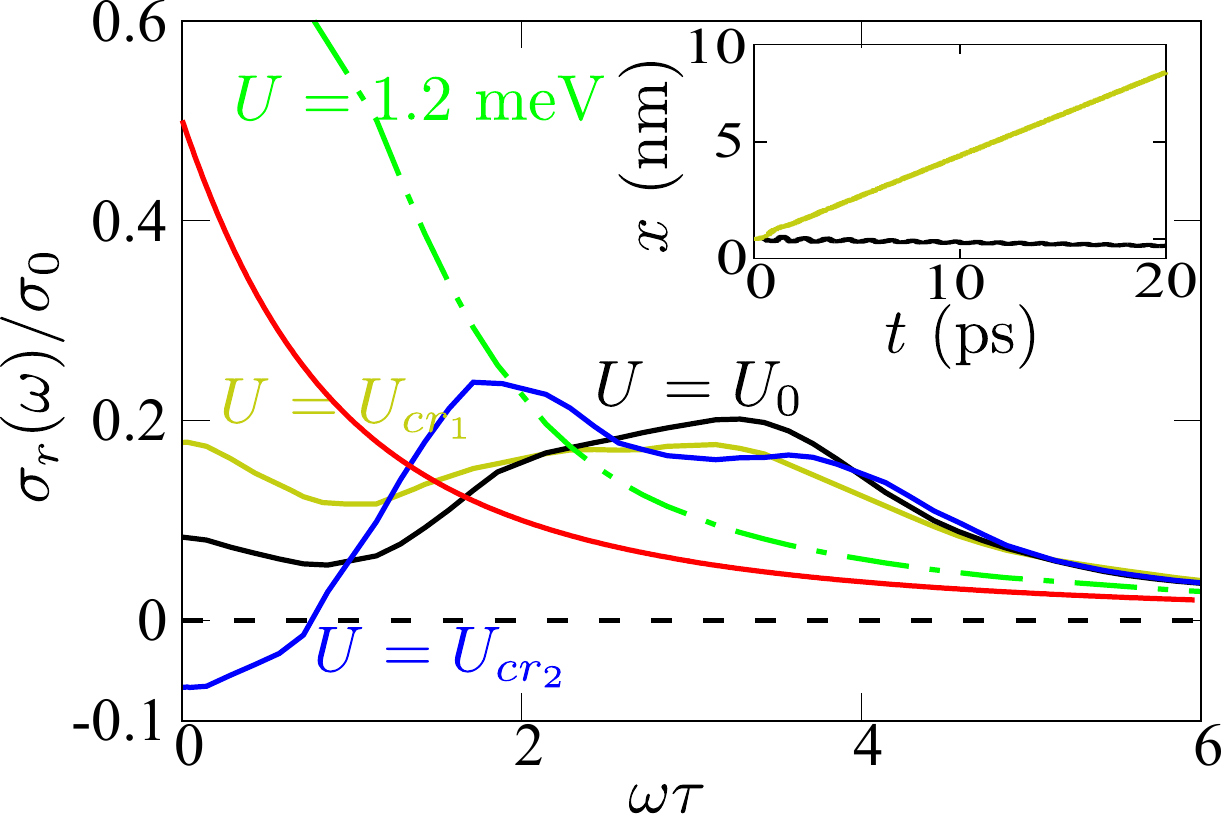}
   \caption{Absorption and gain profiles, $\sigma_r(\omega)$,  at fixed $\omega_s=0.1\tau^{-1}$ and different values of wave  amplitude  ($U$=1.2 meV, $U_{cr_1}=9.55$ meV, $U_0=10$ meV, and  $U_{cr_2}=10.46$ meV). The red curve signifies the free-carrier absorption
whereas the horizontal dashed line marks the zero absorption. Inset: Averaged position of ballistic electron trajectories, $x_a(t)$, calculated for $U=U_{cr_1}$ (yellow curve) and $U=U_0$ (black curve). The calculations have been performed using the parameters of the SL structure with  $v_0/v_s=35$; See Table \ref{tab:secondtable}. 
}
\label{fig10} 
   \end{figure}

\section{\label{sec:level6}CONCLUSION}

In this work we theoretically studied the physical mechanisms of acousto-electric effects associated with miniband charge transport in semiconductor superlattice. Our analysis reveals that superluminal effects, previously described within the framework of Ginzburg-Frank-Tamm theory for electro-magnetic radiation \cite{ginzburg1947doppler,Ginzburg60-UFN, Tamm60}, play an important role in acoustically (strain) driven electron transport in minibands.  In particularly,  our results suggest the realization of superluminal anomalous Doppler and backward normal Doppler effects which are related to the development of dynamical instabilities for certain magnitudes of acoustic stimuli propagating through a superlattice structure.  A kinetic model, based on a semiclassical nonperturbative approach demonstrates that  these effects induced by coherent phonons manifest themselves in excitation of complex Bloch oscillations and formation of electron bunches which can counter-propagate or be localized with respect to the propagating deformation pulse. Such character of electron transport is reflected in characteristic kinks in the dependence  of electron drift velocity upon the acoustic wave amplitude, and can even lead to the  absolute negative mobility. Remarkably,  our calculations indicate that the anomalous Doppler effect for supersonic miniband electrons (effectively superluminal) enables realization of a highly tunable gain similar to the Bloch gain occurring in the voltage driven superlattices \cite{Ktitorov72}. These findings open additional avenues for development of efficient acoustoelectronic devices for microwave and THz ranges by providing alternative means for manipulations of electromagnetic waves in various superlattice electronic devices (SLEDs) \cite{eisele2018high,hramov2014subterahertz} including harmonic multipliers \cite{pereira2017theory,hayton2013phase} and heterostructure millimiter-wave detectors \cite{shao2018fast}. Since semiconductior superlattices serve as building blocks for quantum cascade lasers (QCLs), our theory provides with helpful guidelines for the investigation of unusual acoustoelectric phenomena in QCLs \cite{dunn2020high}, which can be employed for tuning  QCL's  broadband THz emission \cite{opavcak2021frequency}.
\\ In a wider context, similar phenomena can be expected in other miniband systems  subjected to slowly propagating excitations such as ultracold atoms in optical lattices \cite{greenaway2013resonant} or in the physical systems with similar Hamiltonians, e.g. the driven Harper models \cite{kolovsky2012driven}

  \section{Acknowledgments}
 A.A. acknowledges financial support by
the Czech Science Foundation (GAČR) through grant
No. 19-03765. Research of
K.N.A. was partially supported by Marius Jakulis Jason
Foundation. F.V.K. acknowledges financial support from FSU-2021-030/8474000371.
\appendix

\section {\label{App0} Nonlinear dynamics in the rest frame of the acoustic wave}
The system dynamics describing the motion of an electron under the influence of a propagating wave potential is described by the equations (\ref{eq:all-dot}). To conceptualize how the acoustoelectric phenomena discussed in terms of waves and quanta are related to bifurcations mechanisms, we analyze the equations of motion in the moving reference frame  $x'(t)=x(t)+x_i-v_st$. In this frame, the electron is subjected to a time-independent potential $V(x')=-U\sin(k_sx')$ whereas the kinetic energy of an electron is translated into  $\mathcal{E}'(p_x)=\mathcal{E}(p_x)-v_s p_x$. After this  transformation, the Hamiltonian becomes 
\begin{equation}
H'=\mathcal{E}'(p_x)+V(x')
\label{eq:hmf}
\end{equation}
and therefore the equations of motion can be cast as
\begin{subequations}
\label{eq:all-prime}
\begin{eqnarray}
\dot{x}'&=&v_0\sin\left(\frac{p_xd}{\hbar}\right) -v_s, \label{eq:xprime}\\ 
\dot{p}_x&=&k_sU\cos(k_sx'), 
\end{eqnarray}
\end{subequations}
which, in contrast to
Eqs.~(\ref{xdot}), (\ref{pdot}), 
It follows that the equilibria of the dynamical system (\ref{eq:all-prime}) should meet the following conditions
\begin{subequations}
\label{eq:all-fp}
\begin{eqnarray}
v_0\sin \left(\dfrac{p_x d}{\hbar}\right) =v_s,\label{eq:fpx}\\
\cos(k_sx')=0.  \label{eq:fpp}
\end{eqnarray}
\end{subequations}

This dictates the locations of the fixed points 
\begin{subequations}
\label{eq:all-usfp_tildea}
\begin{eqnarray}
x'&=&\frac{\pi}{2 k_s} +\frac{m\pi}{k_s},  \label{eq:unsfp_xtildea}\\
p_x&=&(-1)^{l}\frac{\hbar}{d} \sin^{-1}\left(\frac{v_s}{v_0}\right) +l\frac{\hbar\pi}{d}, \label{eq:unsfp_ptildea}
\end{eqnarray}
\end{subequations}
where $m$ and $l$ are arbitrary integer numbers.
From a viewpoint of a wider miniband ($v_0>v_s$) and therefore higher conduction current, a simple stability analysis reveals that all these fixed points are always either hyperbolic or elliptic points. After dividing (\ref{eq:unsfp_ptildea}) by (\ref{eq:xprime}) and performing integration we obtain the phase trajectory equation 
\begin{eqnarray}\nonumber
x=\dfrac{(-1)^j}{k_s} \sin^{-1} 
 \left\{\sin k_s x_i - \frac{\hbar v_s}{Ud}\left[\frac{v_0}{v_s} \bigg( \cos\dfrac{p_x d}{\hbar}\right. \right. \\- 
 \Biggl. \Biggl. \cos\dfrac{p_i d}{\hbar}\bigg) +\dfrac{p_x d}{\hbar} -\dfrac{p_0 d}{\hbar}\bigg]\bigg\} +j\dfrac{\pi}{k_s}, \label{eq:pp}
 \end{eqnarray}
where $j$ is an integer number, and  ($x_i$,$p_i$) is an initial condition. \\\textit{Cherenkov effects near fixed points.}---
\begin{figure}[h]
   \includegraphics[scale=1]{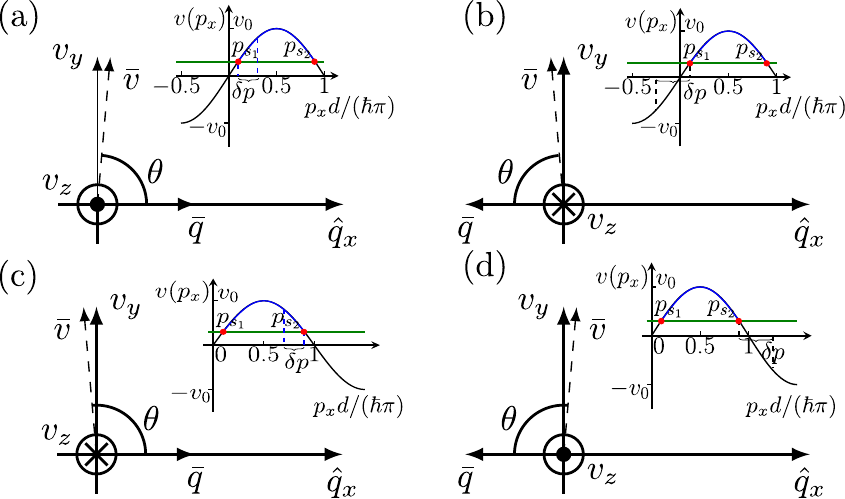}
   \caption{ Schematic representation in the $\bar{q}$--$\bar{v}$ space of the forward  and the backward Cherenkov effects when a charged particle is moving with a supersonic velocity $v(p_x)$ at the center and the borders of the superlattice Brillouin zone. (a) Forward Cherenkov absorption in the proximity of $p_{s_1}$. (b) Forward Cherenkov emission in the proximity of $p_{s_1}$. (c) Backward Cherenkov absorption in the proximity of $p_{s_2}$. (d)  Backward Cherenkov emission in the proximity of $p_{s_2}$. The angle of $\theta$ is the opening angle of the Vavilov-Cherenkov cone indicating an almost perpendicular orientation of the velocity vector $\bar{v}$ with respect to abscissa-located phonon wave vector $\bar{q}$ for $U \ll U_{c1}$. The insets  demonstrate the projection of Cherenkov effects in the active zone described by Fig. \ref{fig2}(a).  
}
\label{fig11} 
   \end{figure}
  We now answer the question whether our  moving particle with velocity $v(p_x)$ larger than $v_s$, is able to induce phenomena analogous to Cherenkov-like conical flow and how the stationary points can help to probe these effects. Strictly speaking, to answer this question one  should then take into consideration  a realistic three-dimensional description by including a quadratic isotropic dispersion law in the Hamiltonian [Eq. (\ref{eq:hamiltonian1})]  for the in-plane kinetic momentum components $p_\parallel=(p_y,p_z)$.  However,  the forward and reversed Cherenkov radiation, in the proximity of the stationary points,  is consistent with the assumption of the 1D emission in Eq. (\ref{eq:absorption}) being finitely small. This implication is easily understood by considering an electron with $v_s/v(\bar{p})\ll  1$ with $\bar{p}$ the total electron momentum. In that case, the Cherenkov mechanism \footnote{This is nothing more than the kinetic description of Cherenkov interactions of sound wave with electrons in metals stemming from Landau absorption mechanism \cite{landau1946vibrations}.} acts mainly in a direction $\hat{q}_x$ almost perpendicular to electron velocity $\bar{v}$ with its x-component being the scalar projection $\bar{v}$ onto  $\bar{q}=q \hat{q}_x$:
\begin{equation}
comp_{\bar{q}} \bar{v}=|v|\cos\theta=\dfrac{q}{|q|}v_0 \sin(p_x),
\end{equation}
\begin{figure}[t]
   \includegraphics[scale=0.45]{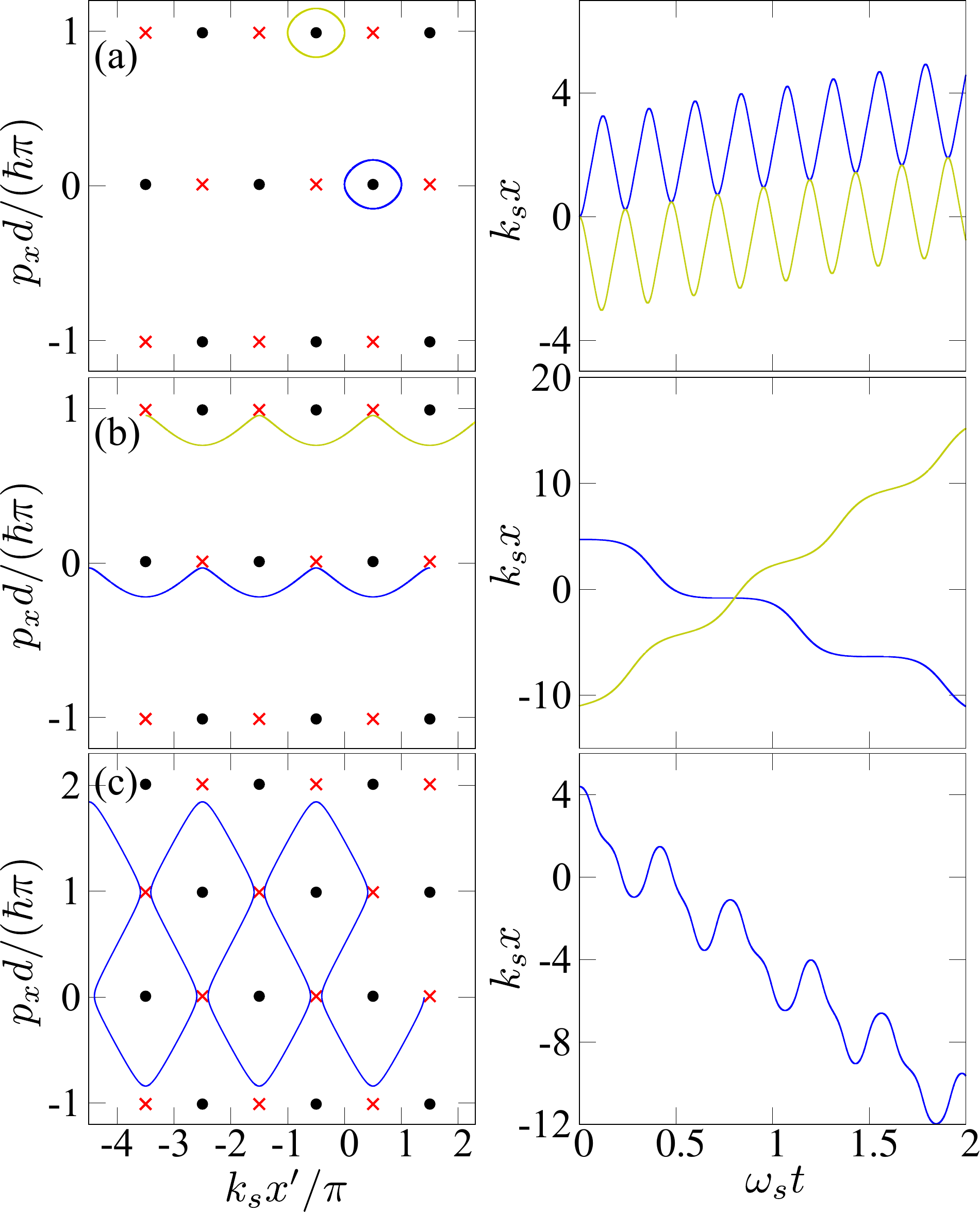}
   \caption{ (a), (b) Phase space trajectories (left panel) and the related electron trajectories (right panel) in real space that can determine the nature of Cherenkov effects in the vicinity of the center and the border Brillouin zone for $U$=1.2 meV. The yellow trajectories correspond to the backward Cherenkov whereas the blue ones to the forward Cherenkov. (c) The complex meandering trajectory and the corresponding trajectory in real space representing the phonon-assisted Bloch oscillation which arises for $U=10$ mev $>U_{cr_1}$. The positions of the elliptic points are depicted by black solid circles and the hyperbolic points by red crosses. The calculations have been performed using the parameters of the SL structure with $v_0/v_s=35$; See Table \ref{tab:secondtable}. }
\label{fig12} 
   \end{figure}
where $\theta$ is the angle between the electron velocity and wavevector $\bar{q}$ of the emitted phonon in the proximity of the stationary points. Hereafter we consider in detail the nature of Cherenkov effects around the center of the first Brillouin zone: (i) when $p_f>$ $p_i=-p_{s_1}$, so that $q>0$. This results in  an  absorption  of phonon  and a positive electron velocity that surpasses sound velocity, which is  forward Cherenkov radiation [Fig. \ref{fig11}(a)]. (ii) When $p_f<$ $p_i=p_{s_1}$, so that $q<0$. This results in  an  emission of phonon  and a negative electron velocity that its norm surpasses sound velocity, which is again forward Cherenkov radiation. Consider now what happens at the border of the first Brillouin zone: (iii) When $p_f<$ $p_i=p_{s_2}$, so that $q<0$. This results in an emission of phonon and a positive electron velocity with $v> v_s$, which is backward Cherenkov. Finally, (iv) when $p_f>$ $p_i=p_{s_2}$ which may lie outside the first BZ (Umklapp process), so that $q>0$ and in this sense  is an absorption of phonon [Fig. \ref{fig11}(d)].
   \begin{figure}[t]
   \includegraphics[scale=0.5]{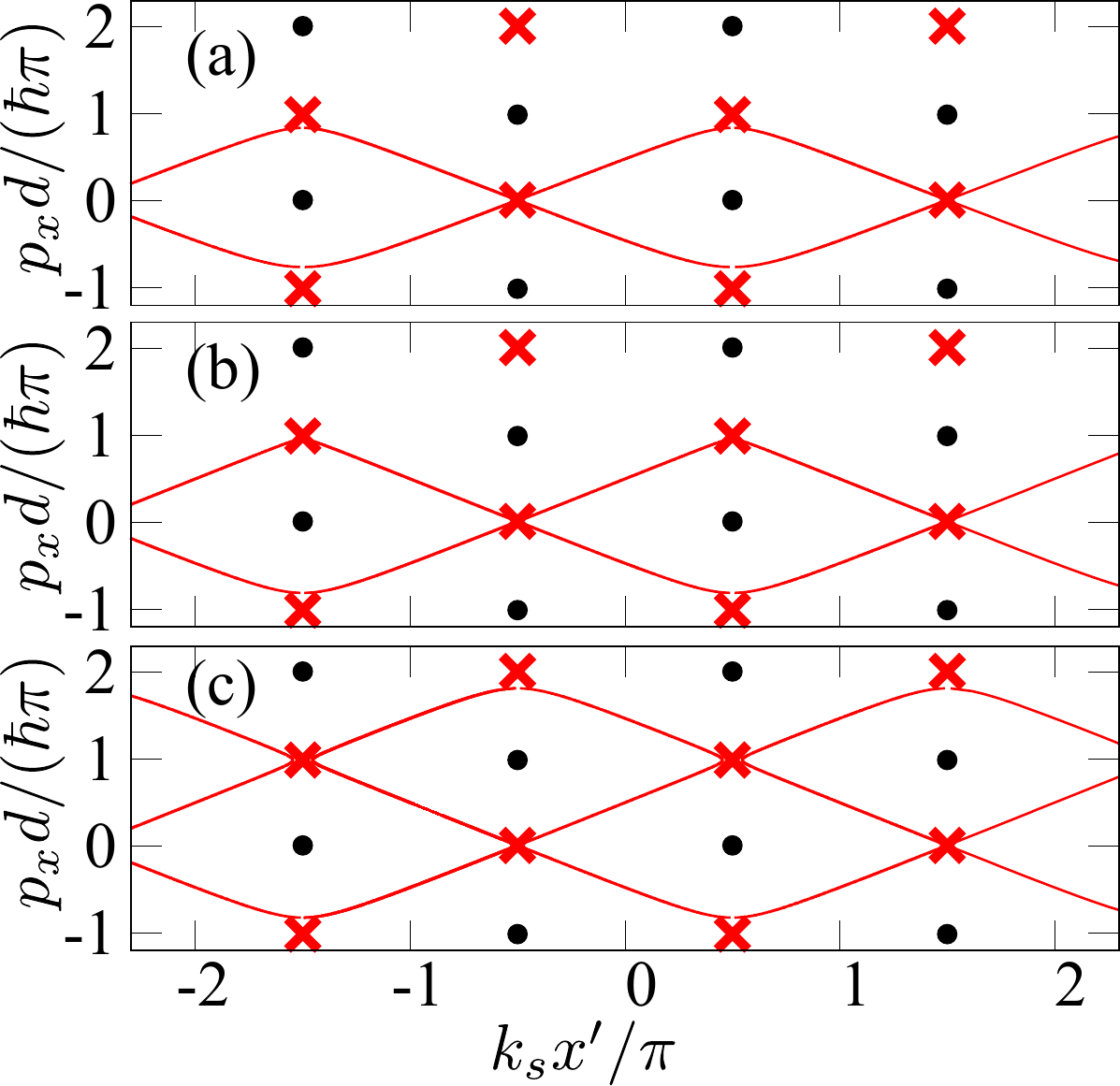}
   \caption{ Separatrix structure in the proximity of the first bifurcation point $U_{cr_1}$ for (a) $U$=9 meV; (b) $U=$9.52 meV; (c) $U$=9.6 meV. The positions of the elliptic points are depicted by black solid circles and the hyperbolic points by red crosses. The calculations have been performed using the parameters of the SL structure with $v_0/v_s=35$; See Table \ref{tab:secondtable}.  }
\label{fig13} 
   \end{figure}
 To find the physically equivalent point inside the first BZ, we have to subtract the module of the reciprocal superlattice vector $|G_0|=2\pi/d$ from $p_f$. Thus, $p_f'=p_f-|G_0|$ will be found in the proximity of $p_{s_3}$ with $v<0$ which results in a backward Cherenkov effect. Practically, we have forward and backward Cherenkov-like conical flow in the center and at the edges of BZ respectively. Similar description one can attain from the elliptic points at the center and edges of the BZ, i.e. normal versus backward Cherenkov. In essence, the counter-wise rotation  of the phase-space trajectories around the elliptic points [left-handed panel of Fig. \ref{fig12}(a)] with  $p_{e_1}=(\hbar/d)\sin^{-1}(v_s/v_0)$ and $p_{e_2}=\pi d/\hbar-(\hbar/d)\sin^{-1}(v_s/v_0)$  designate the nature of the Cherenkov instabilities. Therefore, for  absorption of an infinitely small quasi-particle ($q$) the clockwise trajectory (blue curve) in the proximity of $p_{e_1}$ is related to an electron trajectory  in the real space  that moves  in the positive direction of $x$, $v(p_{e_1}+q)>0$ (forward Cherenkov effect). On the contrary,   the electron trajectory which  moves momentarily in the negative direction,  $v(p_{e_1}+q)<0$ (backward Cherenkov effect), corresponds to the  counterclockwise  trajectory (yellow curve). Those two trajectories are directly associated with the motion of electrons confined by the propagation potential and they coexist for small $U$  with unbounded trajectories [left-handed panel of Fig. \ref{fig12}(b)]  in the moving reference frame. The latter phase-space trajectories are related to unconfined motion [right-handed panel of Fig.\ref{fig12}(b)]  of electrons in real space and their different direction has origin again in the nature of different Cherenkov effects. \begin{figure}[t]
   \includegraphics[scale=0.5]{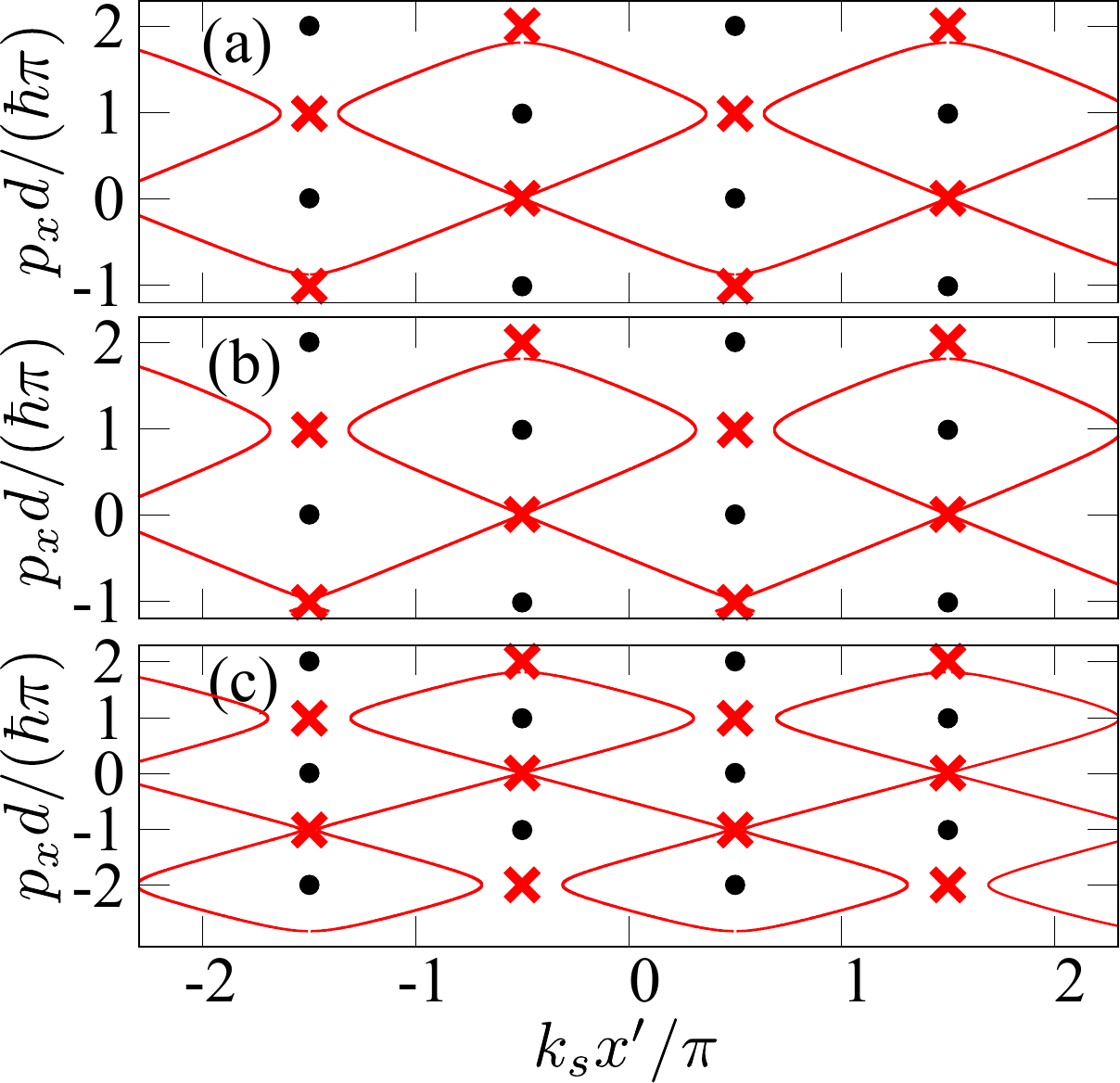}
   \caption{ Separatrix structure in the proximity of the second bifurcation point $U_{cr_2}$ for (a) $U$=10 meV; (b) $U=$10.4 meV; (c) $U$=10.5 meV. The positions of the elliptic points are depicted by black solid circles and the hyperbolic points by red crosses. The calculations have been performed using the parameters of the SL structure with $v_0/v_s=35$; See Table \ref{tab:secondtable}.  }
\label{fig14} 
   \end{figure}\\\textit{Global instabilities (bifurcations).}---  The aforementioned trajectories and the associated Cherenkov effects can well exist both for small and large wave amplitudes. Considering though substantially larger values of $U$ triggers  a series of global instabilities which are attributed to instances  where the manifold of one hyperbolic point touches another hyperbolic point. It has been shown that the conservation of energy, $H'=$ const  [see Eq. (\ref{eq:hmf})], can be used to determine the bifurcation points analytically \cite{apostolakis2017nonlinear}
 \begin{equation}
U_{cr}=\frac{\frac{\Delta}{2}[\cos (p_{s_i}d/\hbar)-\cos (p_{s_j} d/\hbar)]+v_s p_{s_i}-v_s p_{s_j}}{\sin(k_s x_{s_j})-\sin(k_s x_{s_i})}, \label{eq:u_cr-bif}
\end{equation}
where $(x_{s_i},p_{s_i})$, $(x_{s_j},p_{s_j})$ are a set of coordinates of hyperbolic points which are involved in a specific bifurcation. Using these coordinates one can attain the explicit formulations for critical values of the wave amplitude, i.e.  Eq (\ref{eq:u_cr})-(\ref{eq:u_cr-qsp2}).
The bifurcations cause dramatic transformation of the phase space (see Figs. \ref{fig13}, \ref{fig14}) and  result in the emergence of meandering trajectories such as the one depicted in the left-handed panel of Fig.  \ref{fig12}(c) which corresponds to frequency-modulated oscillations in real space  shown in the right-handed panel of Fig. \ref{fig12}(c).   This trajectory in contrast to the ones depicted in Fig. \ref{fig12}(a),(b) follows from a non-conventional acoustoelectric effect which is an effective counterpart of the superluminal anomalous Doppler effect. The discussion under which conditions this phenomenon arises and how relates to global instabilities are presented in detail in Sec. \ref{sec:level2}.

   \section {\label{App2} Ensemble-averaged physical quantities}

By resorting to ensemble-averaged physical quantities in Sec. \ref{sec:level2a} helped us to better understand the role of the Doppler effects in the global transport in the dissipationless limit. In this appendix, we will discuss in detail the proper definition of these quantities starting from a straightforward approach in which we simply average across the initial positions $x_i$ of electrons, or, equivalently  across the initial phases of the acoustic wave
\begin{subequations}
\label{eq:ens-all}
\begin{eqnarray}
v_a(t)&=&v_0 \left\langle \sin\left\{\dfrac{p_x(t) d}{\hbar}\right\} \right\rangle_{x_i},  \label{eq:ens-v}\\
E_a(t)&=&\dfrac{\Delta}{2} \left\langle 1-\cos\left\{\dfrac{p_x(t) d}{\hbar}\right\}  \right\rangle_{x_i}, \label{eq:ens-e}\\
V_a(t)&=&-U \left\langle \sin(k_s x-\omega_s t) \right\rangle_{x_i}, \label{eq:ens-p}
\end{eqnarray}
\end{subequations} 
where   $v_a(t)=\langle \dot{x} \rangle_{x_i}$  is the averaged electron velocity,  $E_a(t)=\langle \mathcal{E}(p_x) \rangle_{x_i}$ is the averaged electron energy and  $V_a(t)=\langle V(x,t) \rangle_{x_i}$ is the averaged potential energy. For example, the value of $\langle v_a(t){x} \rangle_{x_i}$   is determined by averaging $\dot{x}$ over an ensemble of electron trajectories with different $x_i$ from the interval $[-\lambda/2,\lambda/2)$. Here, $\lambda=2\pi/k_s$ designates the space period of the propagating wave.   We are ready now to define the dissipationless approach ($\tau \rightarrow \infty$) for the evaluation of (\ref{eq:Boltz_vd}). Thus, one can formally calculate  the time-averaged velocity $v_m=\langle v_a(t)\rangle_{\Delta t}$ by using

\begin{equation}
v_m=\frac{1}{\lambda}\int_{-\lambda/2}^{\lambda/2}d{x_i} \int_0^{\Delta t} v(t,t_i)\frac{dt}{\Delta t}.
\label{eq:v_m}
\end{equation} 
and the initial conditions
\begin{equation}
x(t_i,t_i)=x_i, \hspace{0.3cm} p_x(t_i,t_i)=0.
\label{eq:v_m2}
\end{equation} 
In this approach $[v(t,t_i), p_x(t,t_i)]$ plays the role of the Vlasov phase and $t_i$ designates the initial time as described in \cite{skadron1974properties, dupree1966perturbation}
 and references therein. \\ However, this approach described by Eq. (\ref{eq:ens-all}) faces shortcomings when we are seeking to define precisely the averaged momentum $p_a$, i.e. the center of the mass of the electron distribution (CMED), $f(p_x,x,t)$, in the quasimomentum space.
Let us now the reduce the dimensionality of $f(p_x, x, t)$ by integrating over all initial phases (from $-\lambda/2$ to $\lambda/2$) of the acoustic wave 
\begin{equation*}
\tilde{f}(p_x,t)=\langle f(p_x,x,t) \rangle_{x_i},
\label{eq:redbolt}
\end{equation*} 
We may obtain  the $x_i$-averaged quantities [Eq. (\ref{eq:ens-v})-(\ref{eq:ens-p})] that were introduced earlier by averaging against $\tilde{f}$,  $\langle .\rangle_{x_i}= d/(\pi \hbar) \int_{-\pi \hbar/d}^{\pi \hbar/d}(.)\tilde{f}(p_x,t)d{p_x}$. One can easily understand the implications of such a choice for  $\langle p_x \rangle_{x_i}$ by considering the $\tilde{f}(p_x,t)$  centered around a symmetric and narrow peak at $p_x=\pi \hbar/d$. In that case, the expectation value $\langle p_x \rangle_{x_i}=0$ contradicts the real value of $\pi \hbar/d$. 
To predict with precision  the CMED we resort to a circular mean angle $p_{\phi}=p_a d/\hbar$ which is determined by the first trigonometric moment $m_1$ of the distribution \footnote{This method relies on superlattice balance equations \cite{isohatala2012devil} and their connection to rotationally symmetric distributions  in directional statistics \cite{mardia2009directional} } 
\begin{equation}
p_{\phi}=\textrm{arg} \hspace{0.1cm} m_1, \hspace{0.5cm} m_1=\langle \textrm{exp}(ip_x d/\hbar)\rangle_{x_i},
\label{eq:direcstat1}
\end{equation}
while the (circular) variance of the distribution is obtained from absolute value, $|m_1|=A$, of the first trigonometric moment
\begin{equation}
\mathcal{V}=1-A.
\label{eq:direcstat2}
\end{equation}
Here the $\mathcal{V}$ takes values between  0 and 1. The value of the lower bound,  $\mathcal{V}=0$, implies  a Dirac $\delta$--distribution function  centered at mean angle $p_{\phi}$, whereas the value of the upper bound, $\mathcal{V}=1$, indicates a distribution function which has no defined mean. 
To analyze though the effects of electron bunching in Sec. \ref{sec:level2a}, we  shall use $A$  which describes the coherence of the electron distribution, i.e. how concentrated is $\tilde{f}$ around its mean. The averaged velocity and averaged energy can be obtained in terms of the variables $p_\phi$, $A$   
\begin{equation}
v_a=\dfrac{A \Delta d}{2 \hbar} \sin(p_{\phi}), \hspace{0.5cm} E_a=\dfrac{A\Delta}{2}[1-\cos(p_{\phi})].
\label{eq:direcstat3}
\end{equation}
From Eqs. (\ref{eq:ens-all}), (\ref{eq:direcstat3}) one finds that
\begin{subequations}
\label{eq:com-all}
\begin{eqnarray}
A\sin(p_{\phi})=\langle \sin(p_x d/\hbar)\rangle_{x_i},  \label{eq:com-v}\\
A(1-\cos(p_{\phi}))=\langle 1-\cos(p_x d/\hbar)\rangle_{x_i}, 
\end{eqnarray}
\end{subequations}
In fact, our numerical calculations further confirmed that the ensemble average approach [Eq. (\ref{eq:ens-v})] and the averaging against circular distribution [see Eq. (\ref{eq:direcstat3})] of electron velocity are equivalent.

\begin{figure*}
   \includegraphics[scale=0.5]{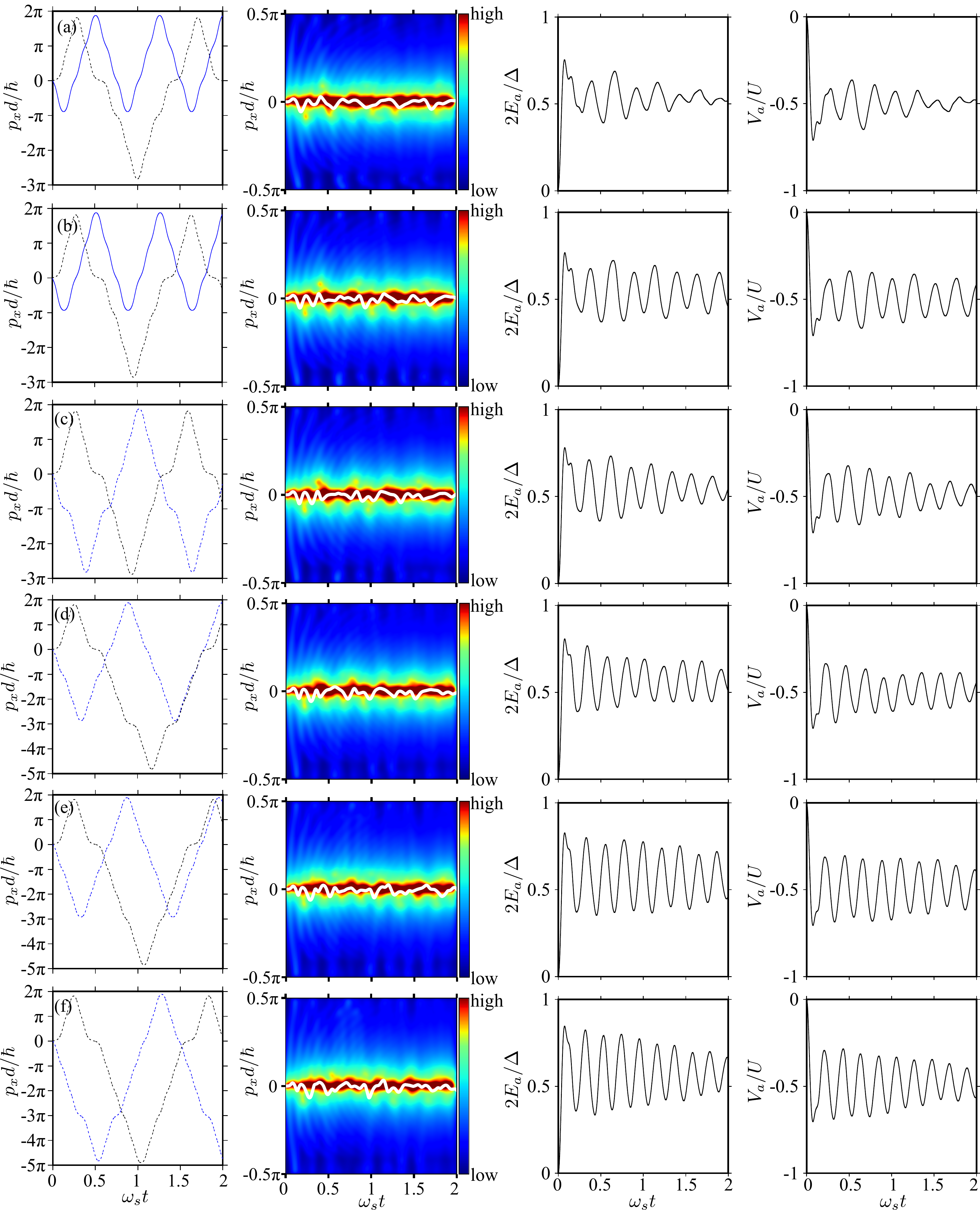}
   \caption{The temporal dynamics of $p_x$--space trajectories subjected to normal and anomalous Doppler instabilities (first panel),  the electron bunches (second panel), the averaged electron energy $E_m$  (third panel) and the averaged potential energy $V_m$ (fourth panel) for  (a) $2U/\Delta=1.06$, (b) $2U/\Delta=1.08$, (c) $2U/\Delta=1.1$ with $U_{c2}<U<U_{cr_3}$ and (d) $2U/\Delta=1.14$, (e) $2U/\Delta=1.17$, (f) $2U/\Delta=1.2$ with $U_{cr_3}<U<U_{cr_4}$.  The black and blue curves in the first panel were calculated for $x_i=-\pi/(2k_s)$ and $x_i=-2/k_s$  respectively. The white lines in the second panel indicate the center of electron distribution as a function of time. The calculations have been performed using the parameters of the SL structure with $v_e/v_s$=22. }
\label{fig15}
\end{figure*}

\subsection*{$p$--space bunching between higher order bifurcations}
 
Figure \ref{fig15} complements Figs. \ref{fig5},\ref{fig6}  by summarizing the properties of wave-like bunching of electrons in momentum space between consecutive higher-order bifurcations. Specifically, we consider three values of $U$ between $U_{cr_2}$ and $U_{cr_3}$, viz., $2U/\Delta$=1.06, $2U/\Delta$=1.08, $2U/\Delta$=1.1  and another three values between $U_{cr_3}$ and $U_{cr_4}$, namely, $2U/\Delta$=1.14, $2U/\Delta$=1.17 and $2U/\Delta$=1.2. When $U>U_{cr_2}$ new type of trajectories  rise, such as the ones depicted by the dashed curves in the first panel of Figs. \ref{fig15}(a)-(c), resulting in nonlinear bunching of the electron distribution function $\tilde{f}$ [see second panel Figs. \ref{fig15}(a)-(c)]. The amplitude of these $p_x$--nonlinear oscillations exceeds the size of the second Brillouin zone caused due to an anomalous Doppler shift and a normal Doppler shift near the end of the first  and the second  Brillouin zone respectively. As $U$ changes between $U_{cr_2}$ and $U_{cr_3}$, the nonlinear bunching becomes more prominent at the value of $2U/\Delta=1.08$ [second panel Fig.\ref{fig15}(c)], coinciding with a local maximum of $v_m(U)$ in Fig. \ref{fig4}. This behavior  implies a balanced mixture of anomalous and normal  Doppler emissions similar to Smith-Purcell effects in photonics crystals \cite{luo2003cerenkov}. Once $U$ lies between $U_{cr_3}$ and $U_{cr_4}$ more complicated trajectories emerge  involving three-phonon  emissions, e.g. black dashed curve in the left panel of the Fig. \ref{fig15}(d). Finally, calculating the averaged electron energy and potential energy further confirms the absorption-emission events  since the  amplitude of their oscillations is enhanced with the variation of $U$, implying the emission (absorption) of larger number of phonons. Note $E_a(t)$ and $V_a(t)$ in third and fourth panel of Figs. \ref{fig15}(a)-(f). 

\section {\label{App1} Derivation of the sound absorption coefficient}

Consider a coherent acoustic wave that propagates through the superlattice. We want to study its evolution and the attenuations effects due to the interaction with the electrons and the crystal under the Hamiltonian (\ref{eq:hamiltonian1}). Then, the derivative of $H$ with respect to time is given by 
\begin{equation}
\dot{H}=\frac{\partial \mathcal{E}}{\partial p_x}\dot{p_x}+\frac{\partial V}{\partial x}\dot{x}+\frac{\partial H}{\partial t}
\label{Hdot}
\end{equation}
By averaging (\ref{Hdot}) over $T_s=2\pi/\omega_s$, we obtain
\begin{equation}
\langle\dot{H}\rangle_{T_s}=2k_sU\left<\left(\dfrac{v_x(p_x)}{v_0}-\dfrac{v_s}{2v_0}\right)\cos(k_s x-\omega_s t)\right >_{T_s}.
\label{Hdot2}
\end{equation}
To calculate the  absorption $\Gamma$ coefficient, we need to substitute the right-hand side of (\ref{Hdot2}) in the definition (\ref{eq:pabs}). The last step requires the exact solution of Boltzmann equation  in the general case by using the path-integral approach [Eq. \ref{eq:Boltz_vd}].

 \section{\label{App5} Small-signal gain in the quasistatic limit}
 \begin{figure}[t]
 \includegraphics[scale=0.58]{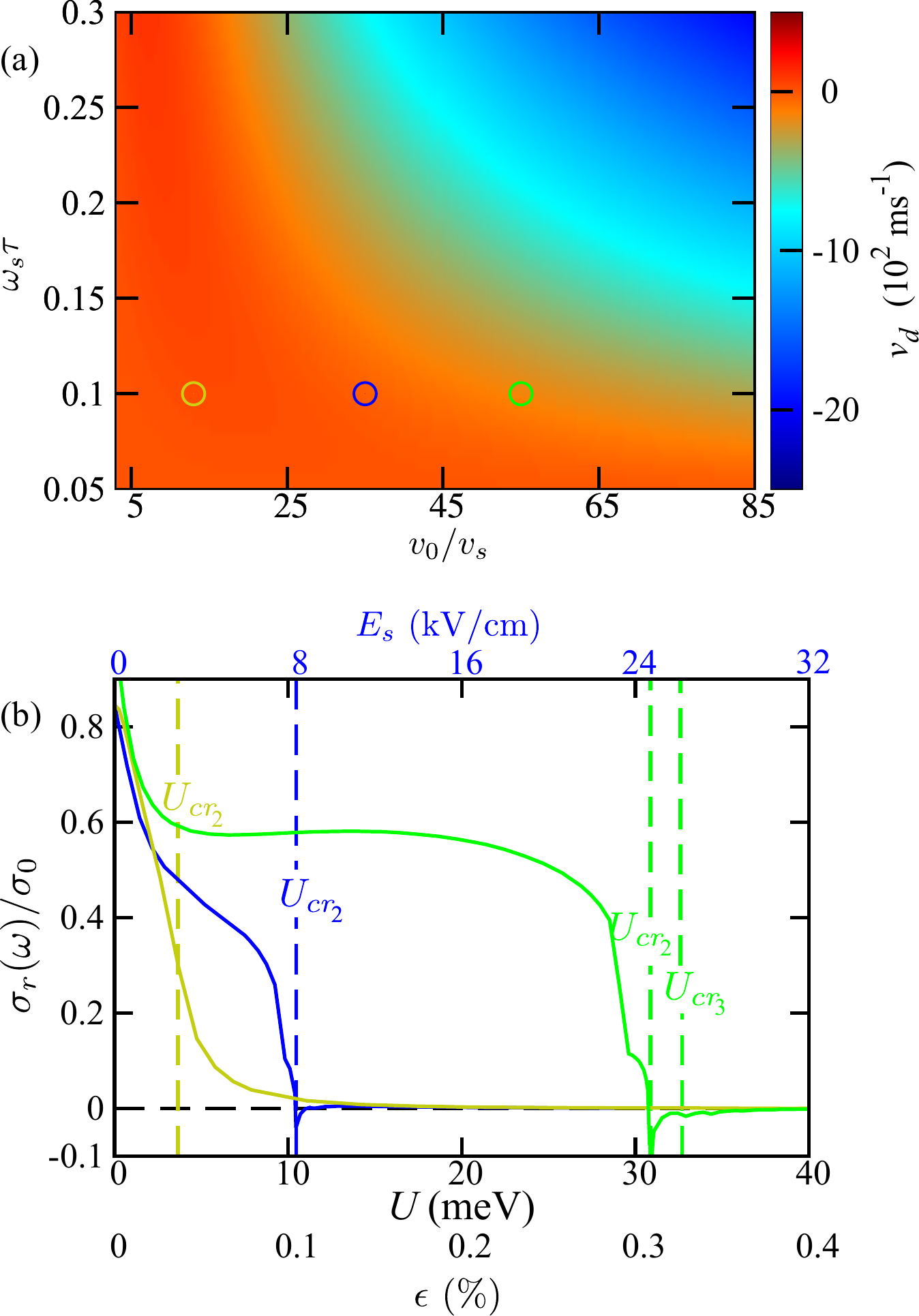}
 \caption{ (a) Color map showing the dependence of the drift velocity $v_d$ on  $v_0/v_s$ and $\omega_s \tau$. The open circles mark the exact values of $v_0$ for which absorption was calculated numerically in the figure below. (b) Absorption $\sigma_r(\omega)$ of the oscillating field $E(t)$ as a function of the wave amplitude $U$ ($1^{\textrm{st}}$  lower scale) or the strain magnitude $\epsilon$ ($2^{\textrm{nd}}$ lower scale) for $\omega\tau<0.1$ and different values of $v_0/v_s=:13, 35, 55$. The upper scale denote the values of the acoustoelectric field [see Eq. (\ref{eq:acoucr})] for a SL structure with $v_0/v_s=35$ and $E_c=2.3$ kV/cm. The vertical lines correspond to the $U=U_{cr_2}$ for different SL parameters; see Table \ref{tab:secondtable}.} 
 \label{fig16}
 \end{figure}
Here we discuss how   $a$--parameter, effectively the product $(v_0/v_s)\times(\omega_s \tau)$, is important for the appearance of  electron bunches with a negative drift  but also whether it  can  influence the interaction of electrons with the electric probe field in the quasistatic limit. By employing again the formulation for the amplitude,  $E_s=k_sU/e$, of a effective acoustoelectric field, we obtain a  generalized condition for the onset of the succesive bifurcations  derived by Eq. (\ref{eq:selectionrules}) which reads \begin{equation}
\left(\dfrac{E_s}{E_c}\right)_n=\alpha+(2n-3)\left(\dfrac{v_s}{v_e}\right)\left(\dfrac{\Delta\hbar}{2\tau}\right)k_s d. 
\label{eq:acoucr}
\end{equation} 
Then from the above equation, it follows  that $\alpha$ can also determine the amplitude of the acoustoelectric field for which the electron localization  [see Eq. (\ref{eq:localcon})]  takes place. In addition,  the second term of Eq. (\ref{eq:acoucr}) depends on $v_s/v_e$  with $v_e=2v_0/\pi$,  which controls the size of the active region [Eq. \ref{eq:deltaE-gen}], the scattering induced broadening $\hbar/\tau$ incorporated in Eq. (\ref{eq:Boltz_vd}) and the product  $k_s d$ which is  associated with  propagation of the acoustic wave. To investigate all the former considerations, we  calculated first the drift velocity $v_d$ as function of $v_0/v_s$ and $\omega_s \tau$ at $U=U_{cr_2}$ [Fig. \ref{fig16}(a)] where the negative drift velocity is possible. The upper-right blue area indicates values ($v_0/v_s$, $\omega_s \tau$)  that result in an enhancement of the backward drift due  to the complex Bloch oscillations which are linked to the anomalous Doppler effect. In contrast, the dark red area designates the region with no reversal of drift velocity to negative values.  As a next step, we calculate  the  absorption, $\sigma_r(\omega,U)$, of an ac probe field as a function of an  arbitrary amplitude, $U$,  of the acoustic wave by resorting to an exact solution of the BTE.   We assume that  frequency of the acoustic wave is  $\omega_s\tau=0.1$ whereas the signal field frequency is much smaller than the inverse relaxation time ($\omega \tau\ll0.1)$. Figure \ref{fig16}(b) demonstrates how the absorption changes with the variation of $U$ for different superlattice parameters and therefore different values of the parameter $a$. By increasing the wave amplitude the incoherent absorption remains positive (yellow curve) for $v_0/v_s$=13 $(a\sim1.3)$ even if $U$ exceeds the critical value $U_{cr_1}$. On the contrary, for a larger miniband width, $\Delta=$ 20 meV  $(a\sim3.5)$, gain is feasible (blue curve) if $U$ reaches the values, where the sign of $\sigma_r$ starts to be sensitive to the Doppler frequency shifts. 
Remarkably, gain exhibits an abrupt change and attains a maximum value close to $U=U_{cr_2}$ similar to the characteristic changes in drift velocity [see Fig. \ref{fig7}]. These effects imply a sensitive dependence  of absorption on the sign of drift velocity and $\alpha$ parameter.  It then follows that the true parameter that determines the mode of absorption of the probe field is $\alpha$ rather than just $\omega_s \tau$. This occurs because $\omega_s$ and $k_s$ enter in the kinetic equations into a combination. 
 Interestingly, for an even larger $\alpha\sim5.5$, the $\sigma_r(\omega)$ dips anew to negative values when $U=U_{cr_3}$ [green curve, Fig. \ref{fig16}(a)].

\section{\label{App5} Key quantities and practical considerations}

\begin{figure}[t]
   \includegraphics[scale=0.9]{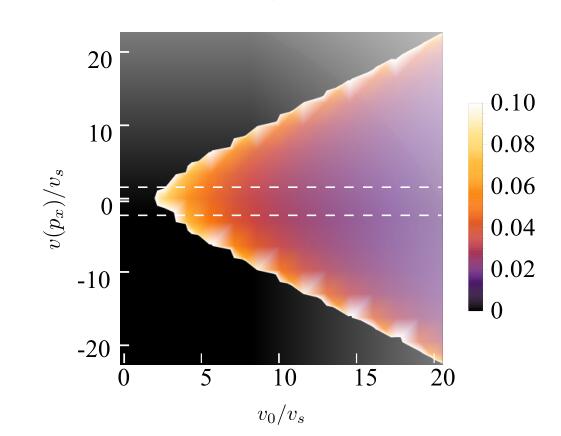}
   \caption{ Colormap showing the probability density $g$ of a particle's velocity $v(p_x)$ within $(-v_0, v_0)$  for $p_x$ uniformly distributed over $(-\pi \hbar/d, \pi \hbar/d)$ and different ratios $v_0/v_s$. 
}
\label{fig17} 
   \end{figure}

\begin{table}[t]
\centering
  \caption{Parameters of superlattices studied}
    \begin{ruledtabular}
   \begin{tabular}{c@{\quad} c@{\quad} c@{\quad} c@{\quad} c@{\quad} c@{\quad} c@{\quad} c@{\quad}}
    $\Delta$ (meV) & $d$ (nm) & $\tau$ (fs) & $v_0/v_s$ & $\sigma_0$ &  \\
    {\tiny miniband width} &{\tiny lattice period} & {\tiny scattering time} & $v_e/v_s$ & {\tiny Drude }  \\ {} &{} & {} & & {\tiny conductivity} \\
    \midrule
    7   & 12.5 & 250 & 13 (10.5) & 5.04 &  \\
    20  & 11.4 & 250  & 35 (22) & 11.98 &  \\
    60  & 6 & 200 & 55 (35) &  7.96 &  \\

  \end{tabular}
  \label{tab:secondtable}
    \end{ruledtabular}
\end{table}
In this appendix we provide further details regarding the superlattices structures considered  to calculate the physical quantities related to acoustoelectric effects. Table \ref{tab:secondtable} outlines the chosen parameters taken from realistic samples \cite{fowler2008semiconductor,shinokita2016strong,Hyart2009} and depicts the critical ratio $v_0/v_s$ that is largely responsible for the demonstration of different types of supersonic behavior and effectively controls the size of the active zone. The entries in the parentheses denote the ratio which is proportional to the effective electron speed ($v_e=2v_0/\pi$) and it is involved in the expressions of the selections rules [Eq. \ref{eq:selectionrules}] allowing us to make a straight-forward comparison with their effects on direct-transport in the quasi-ballistic limit. Furthermore, the fourth column gives the Drude conductivity which is used to scale the dynamical conductivity and therefore complementing Figs.  \ref{fig10}, \ref{fig16}.
\\ One can characterize in a direct way the fraction of supersonic electrons by calculating the probability density $g(v(p_x))$ of miniband electron velocities for the particles starting from  $x_i=0$ and the initial momenta $p_i$ uniformly distributed within the interval $(-\pi \hbar/d, \pi \hbar/d)$, i.e. within the first BZ.
 In this case,  the miniband electron velocities obey a shifted arcsine distribution \cite{arnold1980some} whose probability density function is $g(v)=1/(\pi\sqrt{(v_0)^2-v^2})$ with $-v_0<v(p_x)< v_0$.  Figure \ref{fig17} illustrates a colormap  of $g$ calculated  versus $v(p_x)/v_s$ and $v_0/v_s$. For a typical SL structure with $v_0/v_s$=13,  whose parameters are given in table \ref{tab:secondtable}, the probability density $g$ demonstrates values of small order $\sim 0.01$ in the subsonic (subluminal) region shown by the area between the dashed horizontal lines.  Figure \ref{fig17} reveals that $g$ in the subsonic region  is reduced significantly (transition between yellow and dark-purple colors) with the increase of $v_0/v_s$,  ensuring that $v$ would surpass $v_s$ almost for any $p_x$ at the limit $v_0\gg v_s$.\\
 Table \ref{tab:secondtable} summarizes the critical values associated with the physical processes corresponding to  bifurcations [cf. Eq. \ref{eq:u_cr}]. The bolded entries represent the critical values of $U$ which are all in excellent agreement with numerical simulations, as portrayed by Fig. \ref{fig7}. The wave amplitudes are directly proportional to the deformation potential $D$ and the maximum strain, $\epsilon$, that the acoustic wave creates. For the simulations in this work we consider  $D=10$ eV \cite{fowler2008semiconductor} and  $\epsilon<0.5$ $\%$ which lies within an easily accessible experimental range \cite{fowler2008semiconductor,van2015nonlinear}. These values   are sufficiently large to reach the critical values of $U$ described  in Table \ref{tab:secondtable}.

\begin{table}[H]
\centering
  \caption{Critical values of the wave amplitude for the SL structures in table \ref{tab:secondtable}  }
    \begin{ruledtabular}
   \begin{tabular}{c@{\quad} c@{\quad} c@{\quad} c@{\quad} c@{\quad} c@{\quad} c@{\quad} c@{\quad}}
    & $U_{cr_1}$  & $U_{cr_2}$  & $U_{cr_3}$  & $U_{cr_4}$ &   \\
    {$v_0/v_s$ } &{\tiny meV} & {\tiny meV} &{\tiny meV} & {\tiny meV}  \\ 
    \midrule
   \textbf{13}   & \textbf{3.1} & \textbf{3.92} & \textbf{4.75} & \textbf{5.58} &  \\
    35  &  9.55 & 10.46  &  11.37 & 12.27 &  \\
    55   & 29.14 & 30.87 & 32.59 & 34.31 &  \\
  \end{tabular}
  \label{tab:thirdtable}
    \end{ruledtabular}
\end{table}
\bibliography{draftprx}
\end{document}